\documentclass[12pt,epsf]{article}

\usepackage{verbatim}
\usepackage{anyfontsize}
\usepackage{dsfont}
\usepackage{graphicx}
\usepackage{tikz}
\usepackage{sidecap}
\sidecaptionvpos{figure}{t}
\usepackage{subfigure}
\usepackage{enumitem}
\usepackage[english]{babel}
\usepackage{enumitem}

\usepackage{amssymb,amsmath}
\usepackage{graphicx, xcolor, varwidth}
\usepackage{setspace}
\usepackage[permil]{overpic}
\usepackage{cite}
\usepackage{mathtools}

\DeclareSymbolFont{matha}{OML}{txmi}{m}{it}
\DeclareMathSymbol{v}{\mathord}{matha}{118}

\colorlet{darkblue}{blue!70!black}
\colorlet{darkgreen}{green!70!black}

\usepackage[colorlinks=true,urlcolor=darkblue,linktocpage=true,linkcolor=darkblue,citecolor=darkblue]{hyperref}

\numberwithin{equation}{section}
\DeclareMathSymbol{v}{\mathord}{matha}{118}
\newcommand{\be}{\begin{equation}}
\newcommand{\ee}{\end{equation}}
\newcommand{\bea}{\begin{eqnarray}}
\newcommand{\eea}{\end{eqnarray}}
\newcommand{\bear}{\begin{eqnarray}}
\newcommand{\eear}{\end{eqnarray}}
\newcommand{\beas}{\begin{eqnarray*}}
\newcommand{\p}{\partial}
\newcommand{\eeas}{\end{eqnarray*}}
\newcommand{\ba}{\begin{array}}
\newcommand{\ea}{\end{array}}

\def\ba#1\ea{\begin{align}#1\end{align}}
\def\bs#1\es{\begin{split}#1\end{split}}

\newcommand{\pd}[2][1]{\ifnum#1=1 \frac{\partial}{\partial {#2}} \else
  \frac{\partial^#1}{\partial {#2}^{#1}}\fi}
\newcommand{\dpd}[2][1]{\ifnum#1=1 \dfrac{\partial}{\partial {#2}} \else
  \frac{\partial^#1}{\partial {#2}^{#1}}\fi}
\newcommand{\td}[2][1]{\ifnum#1=1 \frac{d}{d{#2}} \else
  \frac{d^#1}{d{#2}^{#1}}\fi}

\newcommand{\x}{\xi}

\renewcommand{\(}{\left(}
\renewcommand{\)}{\right)}

\newcommand{\nbox}{{\,\lower0.9pt\vbox{\hrule \hbox{\vrule height 0.2 cm \hskip 0.19 cm \vrule height 0.2 cm}\hrule}\,}}

\newcommand{\ie}{{\it i.e.,}\ }

\def\O{{\cal O}}

\newcommand{\bz}{\bar{z}}

\newcommand{\N}{{\cal N}}


\begin{document}
\begin{spacing}{1.3}
\begin{titlepage}

\begin{center}
\vspace*{6mm}

{\Large 
Closed Strings and Weak Gravity from Higher-Spin Causality

}

\vspace*{6mm}

Jared Kaplan and Sandipan Kundu

\vspace*{6mm}

\textit{Department of Physics and Astronomy\\
 Johns Hopkins University\\
Baltimore, Maryland, USA\\}

\vspace{6mm}

{\tt \small jaredk@jhu.edu, kundu@jhu.edu}

\vspace*{6mm}
\end{center}

\begin{abstract}

We combine old and new quantum field theoretic arguments to show that any theory of stable or metastable higher spin particles can be coupled to gravity only when the gravity sector has a stringy structure. Metastable higher spin particles, free or interacting, cannot couple to gravity while preserving causality unless there exist higher spin states in the gravitational sector much below the Planck scale $M_{\rm pl}$. We obtain an upper  bound on the mass $\Lambda_{\rm gr}$ of the lightest higher spin particle in the gravity sector in terms of quantities in the non-gravitational sector. We invoke the CKSZ uniqueness theorem to argue that any weakly coupled UV completion of such a theory must have a gravity sector containing infinite towers of asymptotically  parallel, equispaced, and linear Regge trajectories. Consequently, gravitational four-point scattering amplitudes must coincide with the closed string four-point amplitude for $s,t\gg1$, identifying $\Lambda_{\rm gr}$ as the string scale.  Our bound also implies that all metastable higher spin particles in 4d with masses $m\ll \Lambda_{\rm gr}$ must satisfy a weak gravity condition.

\end{abstract}

\end{titlepage}
\end{spacing}

\vskip 1cm
\setcounter{tocdepth}{2}  
\tableofcontents

\begin{spacing}{1.3}

\section{Introduction}
\label{sec:Introduction}

Advocates have long  argued  on both physical and esthetic grounds that string theory should be our leading candidate for a theory of quantum gravity.   Several recent works  support this viewpoint using arguments based on causality \cite{Hofman:2008ar,Hofman:2009ug,Camanho:2014apa,DAppollonio:2015fly,Hartman:2015lfa,Afkhami-Jeddi:2016ntf,Afkhami-Jeddi:2017rmx,Afkhami-Jeddi:2018own,Afkhami-Jeddi:2018apj,Kaplan:2019soo,Kologlu:2019bco,Belin:2019mnx}.  A rather different theorem from Caron-Huot, Komargodski, Sever, and Zhiboedov (CKSZ) \cite{Caron-Huot:2016icg} shows that perturbative interactions involving higher spin particles\footnote{Throughout the paper by higher spin (HS) particle we always mean a particle with spin $J>2$. For a pedagogical review of HS particles see \cite{Rahman:2013sta, Rahman:2015pzl}.} necessarily organize into asymptotically linear parallel Regge trajectories. An old argument \cite{Amati:1988tn} (see also \cite{Camanho:2014apa}) demonstrates that the resulting amplitudes have imaginary pieces indicative of the production of  long strings.

In fact, a recent bound on the gravitational interactions of massive higher spin (HS) particles \cite{Afkhami-Jeddi:2018apj} implies that a theory with a finite number of  elementary massive HS particles cannot be causal unless there exist  HS states in the gravity sector much below the Planck scale. In this work we will  extend the causality constraints of  \cite{Afkhami-Jeddi:2018apj} and combine them with the theorem of CKSZ to 
 argue that when a theory of metastable HS particles is coupled to gravity, a weakly coupled UV completion of the resulting theory must have HS particles in the gravity sector with many of the properties of fundamental strings. In particular, we will show that any weakly coupled UV completion with a consistent S-matrix can preserve causality at comparatively low energies if and only if {\it gravitational} four-point scattering amplitudes coincide with the tree-level closed string four-point amplitude at high energies. Moreover, we prove that a QFT approximation can exist only for HS particles that obey a {\it weak gravity condition}.

By itself, the CKSZ theorem only applies  when massive HS states are exchanged in a $2\rightarrow 2$ scattering process.  However, in general a theory may contain higher spin particles that are finely tuned such that they are not exchanged in any $2 \rightarrow 2$ scattering process.\footnote{This can happen naturally, for example, when higher spin particles are charged under some global symmetry.} On the other hand, our argument parallels \cite{Afkhami-Jeddi:2018apj}, implying that any theory with even one  elementary  HS particle (of mass $m_{J}$) cannot be causal unless there exist HS states in the {\it gravity sector} at or below $m_{J}$. Then by invoking the CKSZ theorem we conclude that weakly coupled UV completions of such a theory must have a gravity sector including stringy states at high energies, with  the string scale at or below $m_{J}$. We will discuss this  in section \ref{sec:MainScatteringArgument}, but our main focus will be the case of metastable HS particles.

\begin{figure}
\begin{center}
\includegraphics[width=0.6\textwidth]{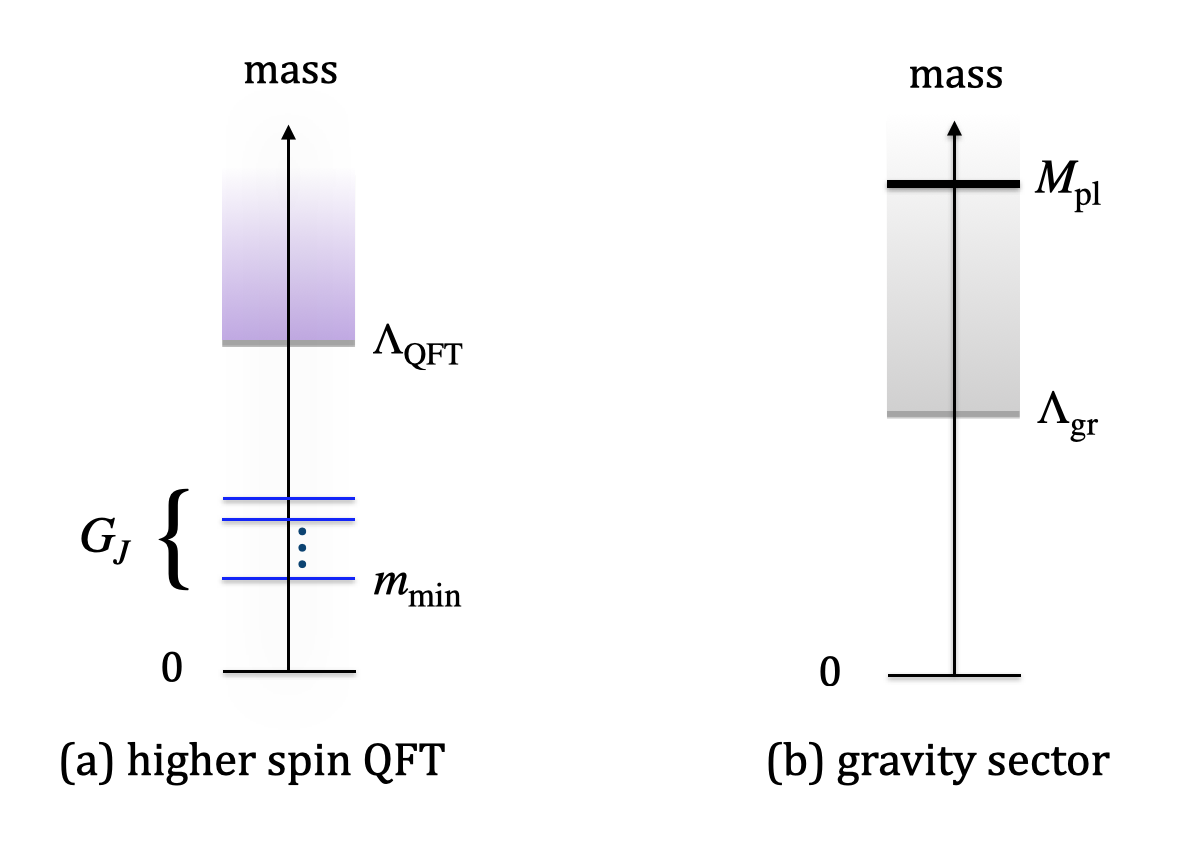}
\end{center}
\caption{ \small  Spectrum of particles with spin $J\ge 3$. Figure (a) represents a QFT that contains higher spin particles $\{G_J\}$ which are effectively elementary below the cut-off scale $\Lambda_{\rm QFT}$. We show that such a theory cannot be coupled to gravity while preserving causality unless there are higher spin states in the gravity sector at $\Lambda_{\rm gr}$ much below the Planck scale, $\Lambda_{\rm gr}\ll M_{\rm pl}$. \label{fig:hs}}
\end{figure}

\subsection{Theories of Metastable Higher Spin Particles}

{It is our goal to establish that a weak gravity condition and stringy states in the gravity sector emerge naturally when one couples theories of metastable HS particles to gravity. So, first we define exactly what we mean by theories of metastable HS particles and the `gravity sector'.}

First, consider a non-gravitational QFT which may contain both low spin and HS particles $\{G_J\}$. We assume that this theory has a consistent S-matrix. This immediately implies that all particles in this theory with spin $J\ge 3$ must be massive \cite{Weinberg:1980kq, Porrati:2008rm}. Furthermore, we assume that the particles $\{G_J\}$ are {\it approximately elementary}  from an effective field theory perspective (see figure \ref{fig:hs} for a pictorial depiction). This can be restated in the following way.  Below some cut-off scale $\Lambda_{\rm QFT}\gg m_J$, where $m_J$'s are the masses of $G_J$ particles, 

\begin{itemize}
\item[] (i)  particles $\{G_J\}$ represent all the degrees of freedom of the theory, 
\item[] (ii)  $G_J$ particles are metastable, so all effective low-energy couplings between low-energy particles are small. 
\end{itemize}
For example, all three-point interactions  
\be\label{eq:interaction}
\langle G_J G_{J'}G_{J''}\rangle \sim \lambda
\ee
must be suppressed: $|\lambda| \ll 1$.\footnote{We also assume that the kinetic mixings between different $G_J$-particles are small.} Thus, the S-matrix in this theory is a meromorphic function with simple poles only at the location of $\{G_J\}$ particles. Of course,  the S-matrix must also be consistent with unitarity, causality, and crossing symmetry.  

\begin{figure}
\begin{center}
\includegraphics[width=0.5\textwidth]{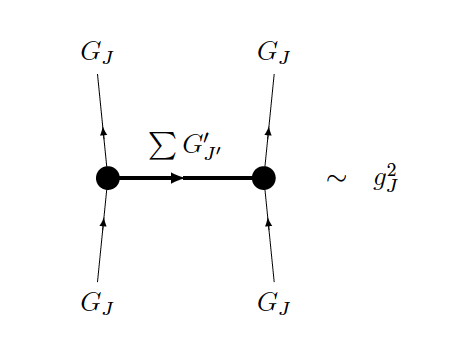}
\end{center}
\caption{ \small  A $G_J$ particle can interact with itself by exchanging one or multiple particles from the $\{G_J\}$-sector. The strength of this interaction is controlled by the parameter $g_J$.\label{intro:int}}
\end{figure}

So, particles in the $\{G_J\}$-sector can be free or weakly interacting below the energy scale  $\Lambda_{\rm QFT}$. In general, a $G_J$ particle can interact with itself by exchanging one or multiple particles from the $\{G_J\}$-sector, as shown in figure \ref{intro:int}. 
We parametrize the strength of this interaction by $g_J\sim \lambda$. This dimensionless parameter will play an important role when we include gravity.

We will eventually couple the above theory to a gravity sector. We assume that the particles $\{G_J\}$ remain effectively elementary  below the energy scale  $\Lambda_{\rm QFT}$ even when we couple the theory to gravity. Hence, the S-matrix in the resulting theory still is a meromorphic function with simple poles which are now located at the position of $\{G_J\}$ particles, the graviton $h_{\mu\nu}$, and other particles in the gravity sector (if any). A $G_J$ particle can decay into two gravitons, however, we impose the following restriction. 
\begin{itemize}
\item[] (iii)  $G_J$ particles are gravitationally metastable: the three-point interactions  
\be\label{eq:decay}
\langle G_J h h \rangle \sim \frac{\lambda_G}{M_{\rm pl}^2}
\ee
are not too large $\lambda_G \sim \O(1)$.\footnote{This condition is violated by bound states in gauge theories at large $N$. In certain situations this assumption can be relaxed. We will discuss this in more detail in section  \ref{sec:MainScatteringArgument}. }
\end{itemize}

We will distinguish between two classes of consistent theories with such HS particles. (A) A theory with finite number of HS particles which are weakly coupled, unitary, and causal up to the energy scale $\Lambda_{\rm QFT} \gg m_J$. For example, massive free HS particles belong in this class. (B) A theory with an  infinite number of HS particles with unbounded spin. This scenario necessarily requires $\Lambda_{\rm QFT}=\infty$, since a thermodynamically healthy theory should not have an accumulation point in the spectrum. Tree-level open string theory belongs in this class. Clearly, there is some overlap between class A and class B theories. If we integrate out all states in a class B theory above some $\Lambda_{\rm QFT}$, we obtain an effective field theory of HS particles which is in class A. However, not all class A theories come from class B theories. For example, a class A theory can come from a strongly coupled UV complete theory.

We  assume that the gravitational sector is consistent on its own, and may contain other particles. To be precise, let us consider  a  $G_J G_J\rightarrow G_J G_J$ scattering for an arbitrary $G_J$. In the limit $M_{\rm pl}\rightarrow \infty$, the tree-level scattering amplitude is a meromorphic function with simple poles only at the location of $\{G_J\}$ particles. When $M_{\rm pl}$ is large but finite, the same scattering amplitude must develop at least one more simple pole corresponding to the graviton. In addition, the scattering amplitude may develop additional simple poles which only disappear in the strict limit of $M_{\rm pl}\rightarrow \infty$. These extra poles represent other particles in the gravity sector. Next, we will argue that these additional gravitational poles are essential to the preservation of causality.

\subsection{A Weak Gravity Condition from Causality}

We will argue that the $G_J$-particles cannot couple to gravity while preserving causality unless there exist HS states\footnote{These cannot simply be a tower of other $G_J$-particles within the non-gravitational sector.} in the gravitational sector. 
Furthermore, we  derive a  bound on the mass $\Lambda_{\rm gr}$ of the lightest HS particle in the gravity sector in terms of $g_J$, $m_J$, and $M_{\rm pl}$. In particular,  in $(3+1)$-dimensions even a conservative estimate implies that $\Lambda_{\rm gr}$ must be small enough such that  
\begin{align}\label{eq:BoundOnStringScale}
&\Lambda_{\rm gr} \lesssim m_J \(\frac{|g_J| M_{\rm pl}}{ m_J}\)^{\frac{1}{2(J-2)}}  \qquad ~~ |g_J|\gtrsim \frac{m_J}{M_{\rm pl}}\nonumber\\
&\Lambda_{\rm gr} \lesssim m_J\qquad ~~~~~~~~~~~~~~~~~~~~~~~~|g_J|\lesssim \frac{m_J}{M_{\rm pl}}\ ,
\end{align}
for all $J\ge 3$ particles. Hence, in general there can be a parametric separation between $\Lambda_{\rm gr}$ and  $m_{\rm min}$. The optimal bound is obtained for the particle in the $\{G_J\}$-sector that minimizes the right hand side of (\ref{eq:BoundOnStringScale}). Of course, the above bound is only a necessary condition but it may be far from being sufficient. For example, for any theory of finite number of elementary HS particles (equivalently type A with $\Lambda_{\rm QFT}=\infty$) causality requires $\Lambda_{\rm gr}\lesssim m_{\rm min}$ even if all HS particles satisfy $|g_J|\gtrsim \frac{m_J}{M_{\rm pl}}$, where $m_{\rm min}$ is the mass of the lightest HS particle in the $\{G_J\}$-sector.

The above bound is closely related to the weak gravity conjecture \cite{Vafa:2005ui,ArkaniHamed:2006dz} (for a recent review see \cite{Palti:2019pca}). It is also reminiscent of the higher spin swampland conjecture of \cite{Klaewer:2018yxi}. Consider a theory of stable or metastable HS particles coupled to gravity in 4d. The theory, as we stated before, must contain gravitational states at or below $\Lambda_{\rm gr}$. However, we can still obtain a low energy QFT description for a set of light HS particles by integrating out states above $\Lambda_{\rm gr}$. Hence, a QFT description exists for a HS particle of mass $m_J$ and interaction strength $g_J$ only when $\Lambda_{\rm gr}\gg m_J$. The bound (\ref{eq:BoundOnStringScale}) implies that the QFT approximation necessarily breaks down when $|g_J|\lesssim \frac{m_J}{M_{\rm pl}}$. This is precisely the statement that the non-gravitational interaction between the particles is weaker than the gravitational interaction. We will also argue that such a particle must have stringy scattering amplitudes.

\begin{figure}
\begin{center}
\includegraphics[width=0.4\textwidth]{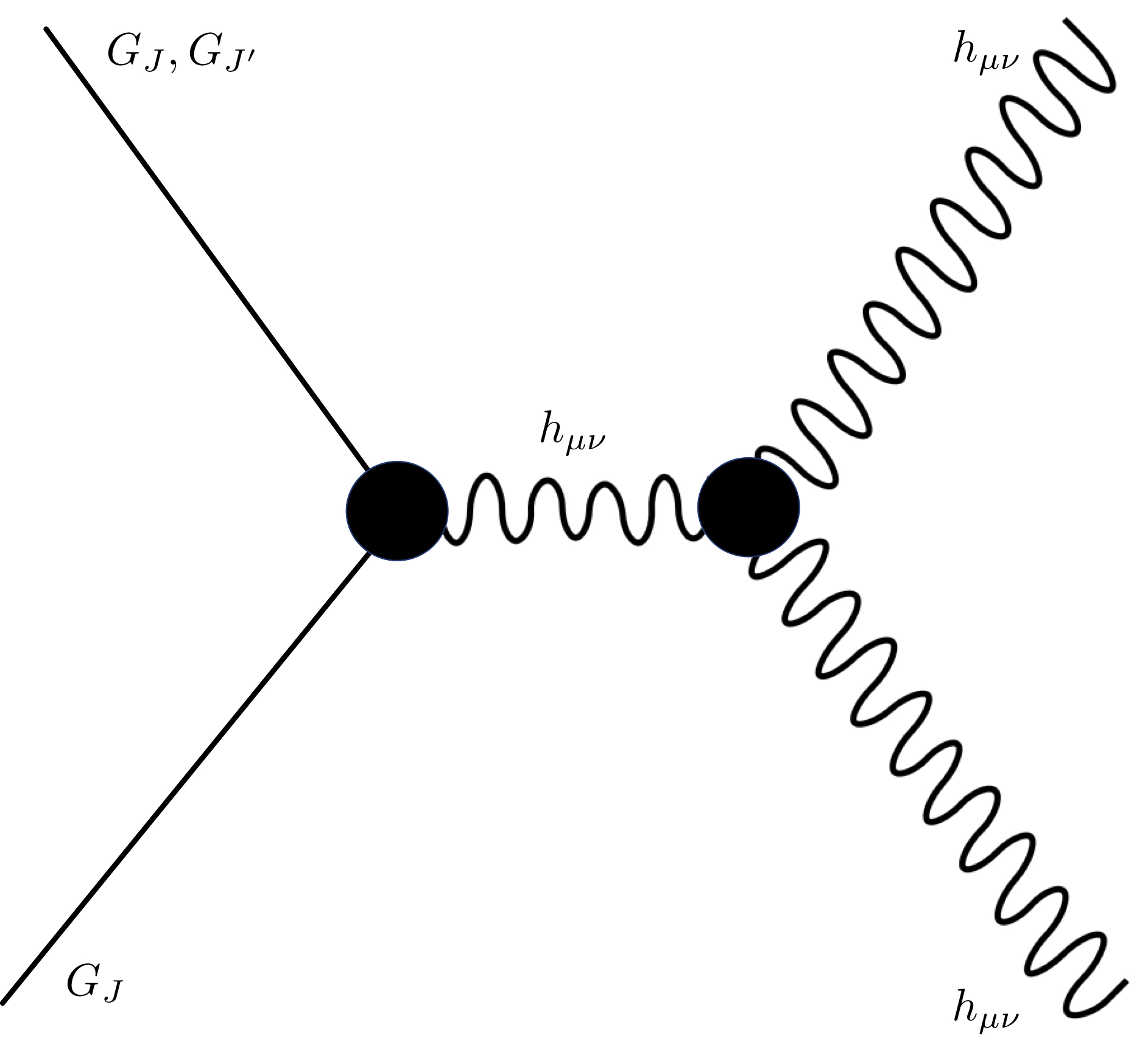}
\ \ \ \ \ \ \ \ \
\includegraphics[width=0.4\textwidth]{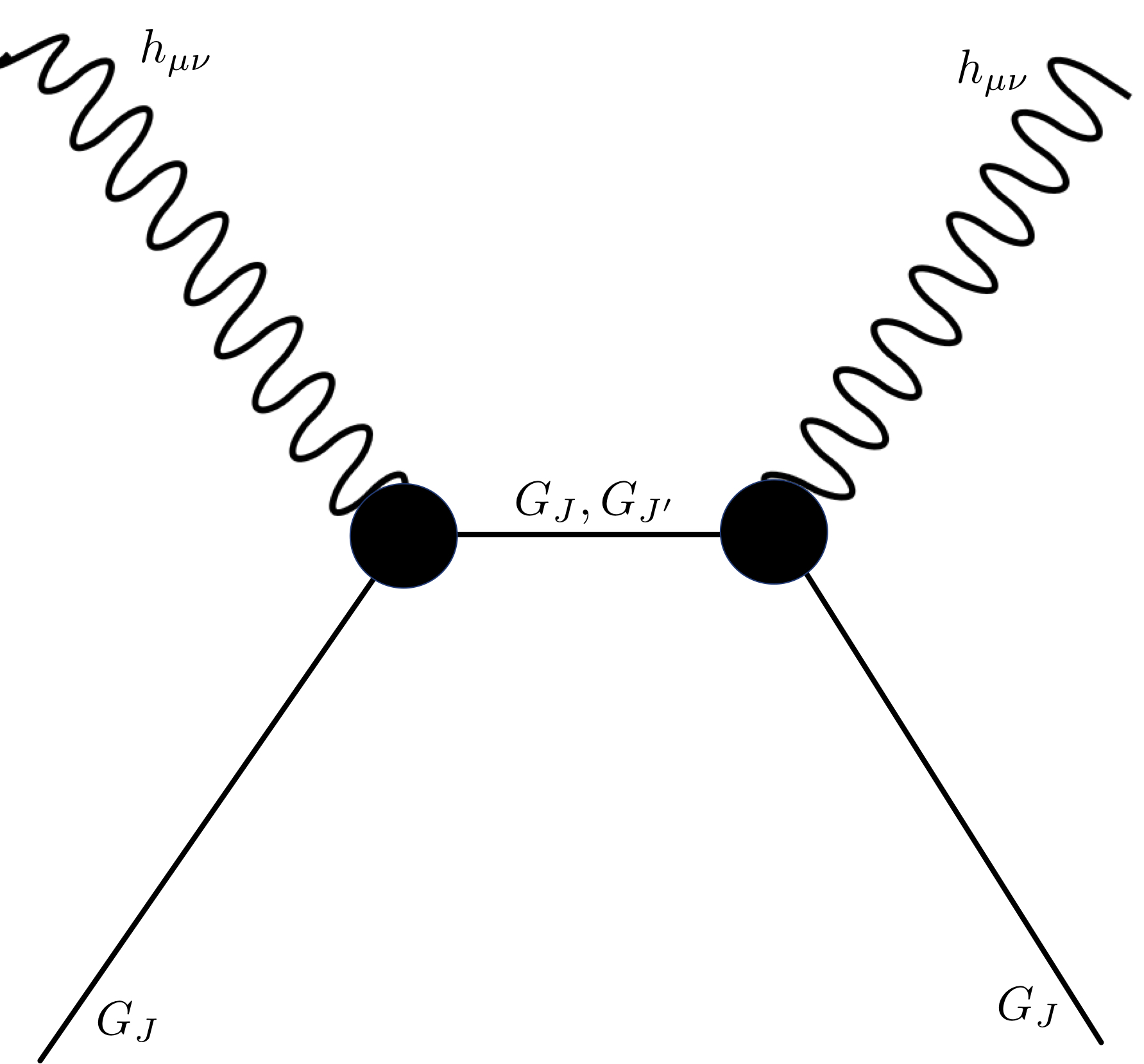}
\end{center}
\caption{\label{fig:BasicDiagrams} \small The process at left constrains the $G_J G_{J'} h$ vertex, requiring it to be suppressed by a power of the scale $\Lambda_{\rm gr}$ where new HS particles contribute to the process.  The process at right involves the exchange of a $J \geq 3$ HS particle $G_J$.  The contributions from other HS particles $G_{J'}$ in the diagram at right are suppressed compared to $G_J$ due to the constraints on the diagram at left.  \label{fig:CausalityConstrainedScatteringDiagrams}}
\end{figure}

\begin{figure}
\begin{center}
\includegraphics[width=0.8\textwidth]{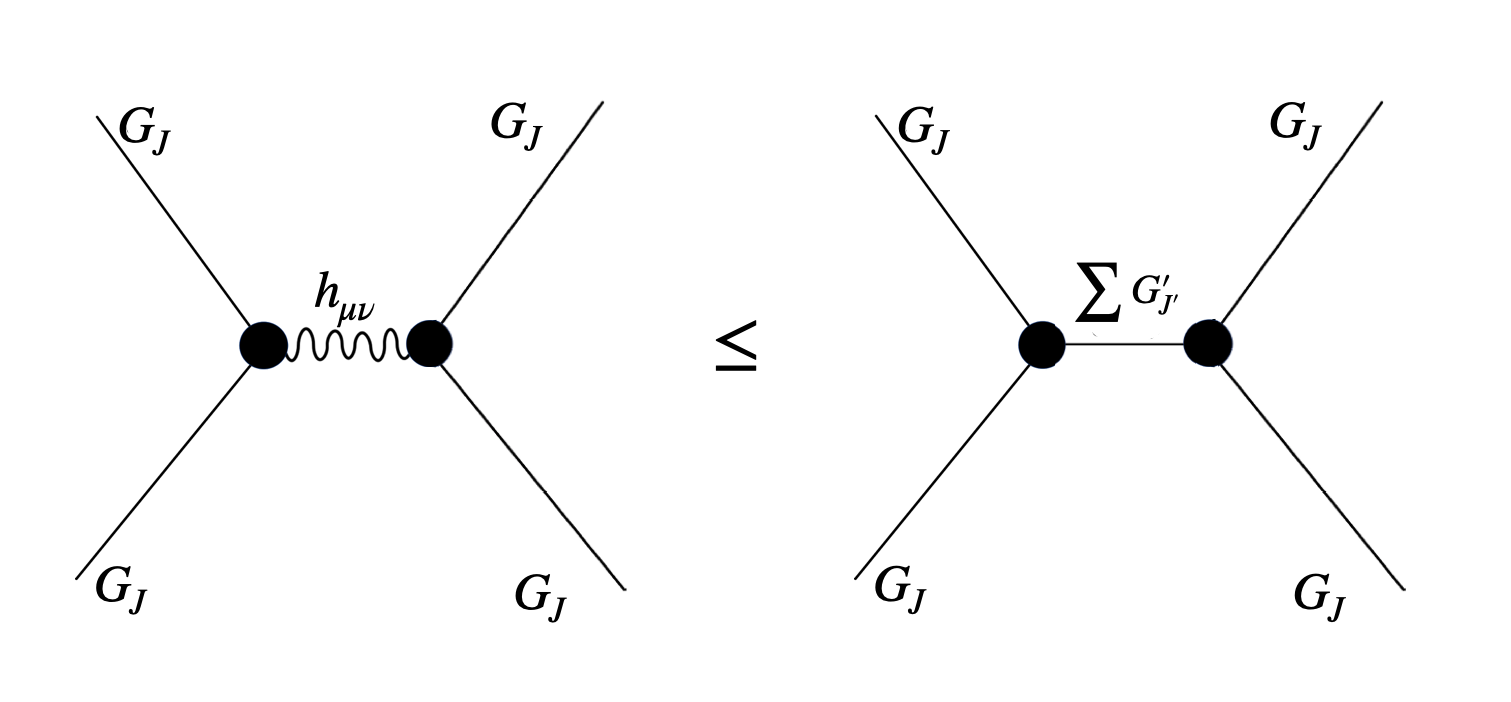}
\end{center}
\caption{ \small Causality imposes a constraint on the $G_J G_J\rightarrow G_J G_J$ scattering amplitude for $J\ge 3$ in 4d. In the impact parameter space, the gravitational part of the scattering amplitude must be smaller than the non-gravitational part for $\Lambda_{\rm gr}\gg \sqrt{s}\gg m_J, \frac{1}{b}$, where $b$ is the impact parameter.    \label{fig:weakgravity}}
\end{figure}

These ideas are motivated by recent work \cite{Afkhami-Jeddi:2018apj} that greatly curtails the existence of stable higher-spin particles $G_J$. We can break our main argument  into three steps, which we work through in detail in section \ref{sec:MainScatteringArgument}.  Here is a brief outline:
\begin{enumerate}
\item  HS particles  $G_J$ have 3-pt couplings to the graviton $h$ that are highly constrained.  There is a unique structure for the $G_J G_J h$ vertex in $D=4$ dimensions that allows a tree-level $G_J h \to G_J h$ scattering to proceed  above the scale of the mass $m_{J}$ of the HS particle without causality violations. Interestingly, this unique structure matches\footnote{We thank Simon Caron-Huot for pointing this out to us.}  the universal coupling \cite{Arkani-Hamed:2019ymq} of gravitons to Kerr black holes in the classical limit.

\item  Vertices involving distinct HS particles $G_{J} G_{J'} h$ must be suppressed by  positive powers of  $1/\Lambda_{\rm gr}$, a new scale $\Lambda_{\rm gr} > m_{J}, m_{J'}$. This follows from causality which implies that the process $G_J h \to G_{J'} h$ is bounded by the process $G_J h \to G_{J} h$.  New HS particles must be present in the gravity sector at $\Lambda_{\rm gr}$.   However, from this argument alone we cannot constrain $\Lambda_{\rm gr}$, and it would be self consistent to simply set the vertex $G_{J} G_{J'} h$ to zero and include no new HS particles. Nevertheless, the fact that all vertices $G_{J} G_{J'} h$ are suppressed will play a crucial role in deriving the bound (\ref{eq:BoundOnStringScale}).

\item \label{thirdstep} Finally, we can complete the argument by studying the scattering  $G_J G_J \rightarrow h h$.  In this step the main physical point is that this process involves the diagram on the right of figure \ref{fig:CausalityConstrainedScatteringDiagrams}. This can be interpreted as an exchange of $G_J$, and so it naively seems to badly violate causality constraints  \cite{Camanho:2014apa} and the chaos bound \cite{Maldacena:2015waa} for $J\ge 3$.\footnote{In \cite{Kologlu:2019bco}, authors proposed that  in any UV complete theory of gravity coincident gravitational shocks should commute. The diagram on the right of figure \ref{fig:CausalityConstrainedScatteringDiagrams} also appears when one studies commutativity of coincident shocks for HS particles coupled to gravity. Indeed, causality constraints obtained from the first two steps of our argument guarantee that coincident gravitational shocks commute below the energy scale $\Lambda_{\rm gr}$. }

However this conclusion is premature for two reasons: (1) the exchange of an infinite tower $\sum G_J$ of other HS particles (in the non-gravitational sector) might resolve the problem, and (2) causality constraints cannot be directly applied because the incoming and outgoing states are different.  The first issue with the argument is resolved by the previous step, which showed that $G_J G_{J'} h$ vertices  must be additionally suppressed by $\Lambda_{\rm gr}$, and so these exchanges cannot rescue causality unless the bound (\ref{eq:BoundOnStringScale}) is satisfied.  The second issue can be resolved by studying a scattering experiment involving a coherent superposition of many states of the form $\alpha G_J + \beta h$.   Now the diagram on the right of figure \ref{fig:CausalityConstrainedScatteringDiagrams} contributes and coherence guarantees that a 2-to-2 scattering process exists.

A careful study of this process imposes a bound on the diagram on the right of figure \ref{fig:CausalityConstrainedScatteringDiagrams}. This, in turn, connects the parameters of the non-gravitational theory to the scale $\Lambda_{\rm gr}$ where new HS particles contribute in the gravity sector, bounding $\Lambda_{\rm gr}$ as in equation (\ref{eq:BoundOnStringScale}). 
\end{enumerate}
This argument implies the existence of HS particles in the gravity sector at a scale $\Lambda_{\rm gr} \ll M_{\rm pl}$.  As a byproduct, we derive that the gravitational scattering amplitude $G_J G_J\rightarrow G_J G_J$ of any HS particle with mass $m_J \ll \Lambda_{\rm gr}$ must be smaller than the non-gravitational scattering amplitude of $G_J G_J\rightarrow G_J G_J$ in the impact parameter space (see figure \ref{fig:weakgravity}). It is possible that the connection between causality and the weak gravity conjecture is more general. In fact, this connection was also observed in a wide class of theories \cite{Hamada:2018dde}. 

The fact that there can be a parametric separation between $\Lambda_{\rm gr}$ and $m_J$ is true only in $D=4$ dimensions. In this sense, 4d is special because it  allows for a field theoretic description of HS theories coupled to gravity. For $D>4$, the three-step argument discussed above is not actually needed. The first step alone implies that a theory with approximately elementary HS states in $D>4$ must have new HS states in the gravity sector at or below $m_{\rm min}$.\footnote{The CKSZ theorem applies here as well implying any such theory in $D\ge 5$ dimensions only have a stringy description. On the other hand, in 3d there are non-string theory models with infinite number of interacting massive higher spin fields \cite{Metsaev:2020gmb,Skvortsov:2020pnk}.}

\subsection{Closed Strings}
Theories of HS particles are known to be highly constrained \cite{Weinberg:1964ew,Weinberg:1980kq,old-weinberg,Ferrara:1992yc, Porrati:1993in,Cucchieri:1994tx,Benincasa:2007xk,Porrati:2008rm,Schuster:2008nh,McGady:2013sga,Camanho:2014apa,Caron-Huot:2016icg,Arkani-Hamed:2017jhn,Afkhami-Jeddi:2018apj,Bellazzini:2019bzh,Melville:2019tdc,Freivogel:2019mtr}. First of all,  HS exchanges must always come in infinite towers with fine-tuned masses and coupling constants. Hence, any theory of approximately elementary HS particles can only couple to a gravity sector that necessarily contains the graviton and an infinite tower of  HS particles above $\Lambda_{\rm gr}$. Assuming the resulting theory is still weakly coupled, the gravitational scattering amplitude  is a meromorphic function that obeys unitarity and crossing symmetry.\footnote{{Note that we are not claiming that all UV completions must be weakly coupled.}} Furthermore, we assume that the (gravity) spectrum does not have any accumulation point (all the assumptions are discussed in detail in section \ref{sec:QCDGravity}). As we explain in section \ref{sec:CKSZ}, we can then invoke  the S-matrix based argument of \cite{Caron-Huot:2016icg} to conclude that the gravity sector must contain  infinitely many asymptotically parallel, equispaced, and linear Regge trajectories. In particular, the gravitational scattering amplitude in the unphysical regime $s,t\gg  \Lambda_{\rm gr}^2$ must coincide with the tree-level Gross-Mende string amplitude \cite{Gross:1987kza}\footnote{The large $s,t$ limit should be taken by avoiding poles. This can be achieved by taking $\mbox{Re}\ [s, t]\rightarrow \infty $ with $\mbox{Im}\ [s, t]>0$.}  
\be\label{amp_asym}
\lim_{s,t\gg  \Lambda_{\rm gr}^2}  A_{\text gravity}(s,t)=A_0 \exp\(\frac{\alpha'}{2}\((s+t)\ln (s+t)-s\ln s-t\ln t\)\)\ ,
\ee
where the Regge slope is given by $\alpha'\approx \frac{1}{\Lambda_{\rm gr}^2}$. It is natural to identify this asymptotic amplitude with the large $s,t$ limit of the tree-level  four-point amplitude of fundamental closed strings. Thus, above $\Lambda_{\rm gr}$ the theory must become stringy, so that in fact $\Lambda_{\rm gr}$ provides an effective string scale
\be
M_{\rm string}\approx\Lambda_{\rm gr}\ .
\ee

\begin{figure}
\begin{center}
\includegraphics[width=0.7\textwidth]{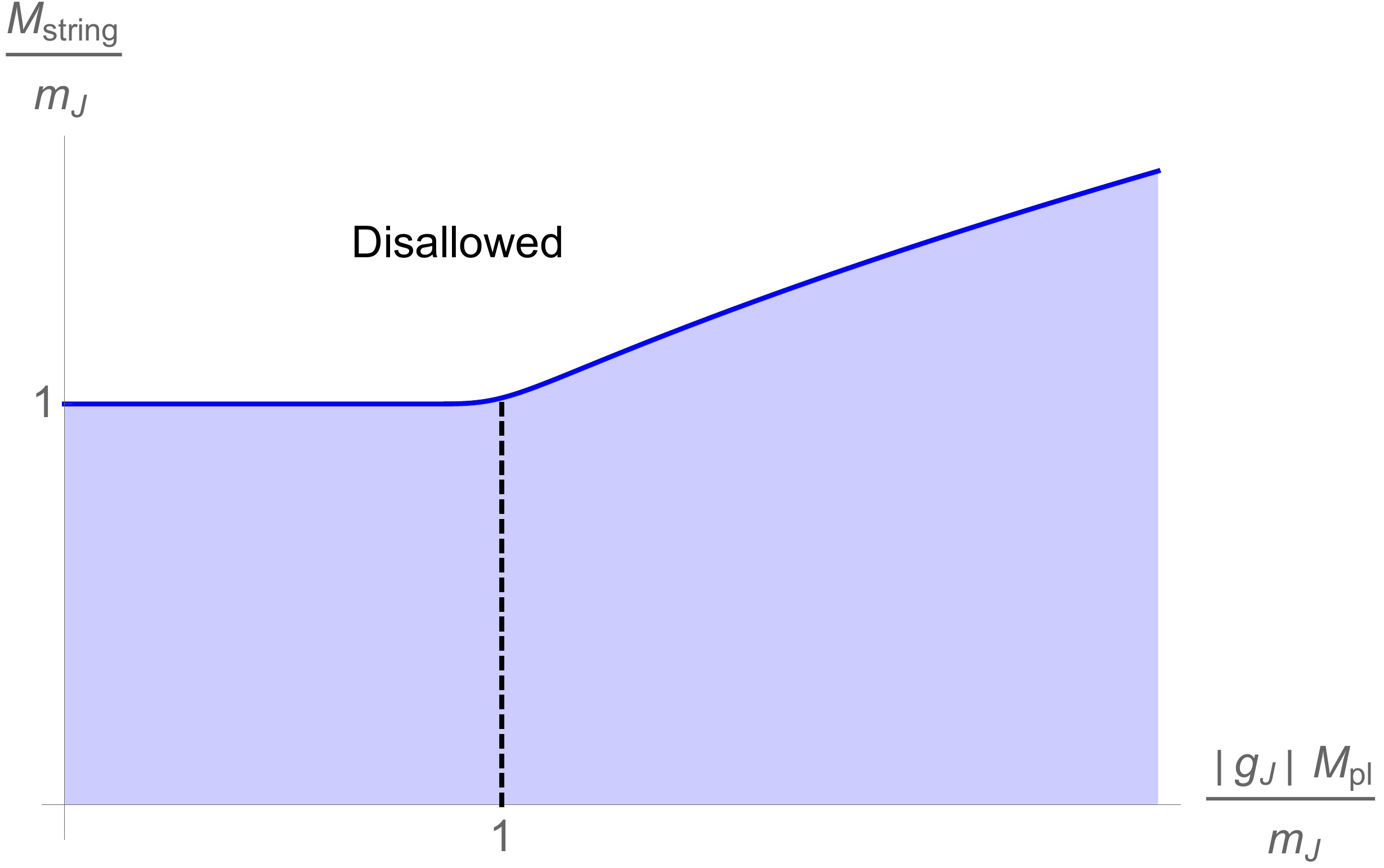}
\end{center}
\caption{ \small  {A schematic exclusion plot for the string scale $M_{\rm string}\approx\Lambda_{\rm gr}$ as a function of $g_J$. In the unshaded region, HS theories violate causality. The solid blue line represents the bound (\ref{eq:BoundOnStringScale}) for $J=3$. For $J>3$, the bound asymptotes to 1 at a faster rate. The weak gravity condition is satisfied in the right of the dashed black line.}\label{fig:intro}} 
\end{figure}

The bound (\ref{eq:BoundOnStringScale}) now has an obvious interpretation as a bound on the string scale.
The  bound on the string scale, as summarized in figure \ref{fig:intro}, appears quite surprising even from an effective field theory viewpoint, since the UV completion of this theory has been constrained in a rather profound way by its IR dynamics.  Certain UV Lagrangians that  seem healthy based on a cursory analysis are in fact inconsistent if we do not include stringy states above $\Lambda_{\rm gr}$.

Confining gauge theories in 4d contain glueballs (and mesons) of spin $J \geq 3$ and lifetime parametrically $\propto N^2$, (and $\propto N$) so that as $N \to \infty$ these higher-spin particles become stable and effectively elementary \cite{tHooft:1973alw, tHooft:1974pnl,Witten:1979kh}. Results of this paper suggest that confining large $N$ gauge theories when coupled to gravity are constrained by causality. {However, these constraints are more subtle since in general gravitational decays of glueballs and mesons are enhanced by $N$, violating (\ref{eq:decay}). We will discuss this separately \cite{Kaplan:2020tdz}.  }

~\\

The rest of the paper is organized as follows. We begin with a review of some basic properties of scattering amplitudes in flat space and list all our assumptions in \ref{sec:QCDGravity}.
  In section \ref{sec:MainScatteringArgument} we present our main argument. Then we briefly review the CKSZ theorem in section \ref{sec:CKSZ} and combine it with the results of section \ref{sec:MainScatteringArgument} to conclude that metastable HS particles can only couple to a gravity sector that has an asymptotically unique UV completion.

\section{Scattering Amplitudes in Flat Space}
\label{sec:QCDGravity}

\subsection{Causality at Low Energies}\label{sub:causality}
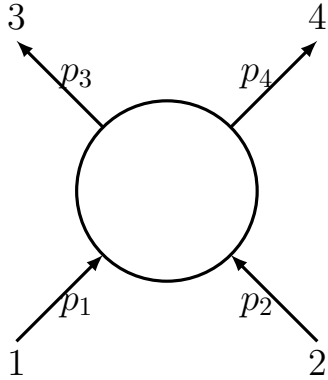
\begin{figure}
\begin{center}
\usetikzlibrary{decorations.markings}    
\usetikzlibrary{decorations.markings}    
\begin{tikzpicture}[baseline=-3pt,scale=0.4]
\begin{scope}[very thick,shift={(4,0)}]
\draw (0,0) circle [radius=3];
\draw[-latex]  (2.12,2.12)--(5,5) ;
\draw[-latex]  (-5,-5)--(-2.12,-2.12);
\draw [-latex] (5,-5)--(2.12,-2.12) ;
\draw [-latex](-2.12,2.12)--(-5,5) ;
\draw(3,3)node[above]{\large $p_4$};
\draw(-3,3)node[above]{\large $p_3$};
\draw(3,-3)node[below]{\large $p_2$};
\draw(-3,-3)node[below]{\large $p_1$};
\draw(5,5)node[above]{\large $4$};
\draw(-5,5)node[above]{\large $3$};
\draw(5,-6.5)node[above]{\large $2$};
\draw(-5,-6.5)node[above]{\large $1$};
\end{scope}
\end{tikzpicture}
\end{center}
\caption{\label{fig:scattering} \small $2\rightarrow 2 $ scattering of particles.}
\end{figure}

We impose the condition that our theory of HS particles can be coupled to gravity without violating causality at energies much below the scale of new physics  $\Lambda_{\rm gr}$. This will be implemented by imposing the following condition. 
\begin{itemize}
\item{The phase shift $\delta(s,\vec{b})$ of eikonal scattering, where $\vec{b}$ is the impact parameter, is non-negative and does not grow faster than $s$.}
\end{itemize}
If $\delta(s,\vec{b})$ grows faster than $s$ for large $s$, it was argued in \cite{Camanho:2014apa} that it can be exploited to send a signal outside the lightcone. On the other hand, even when the eikonal phase shift grows as $\sim s$, it still determines the Shapiro time delay and hence should be positive \cite{Camanho:2014apa}. Before we explain all the subtleties  of this requirement, let us first introduce our notations.  We use the following null coordinates in $\mathbb{R}^{1, 3}$ 
\ba\label{eq:metric}
ds^2  = -du dv + d\vec{x}_{\perp}^2
\ea
in which we consider   a $2\rightarrow 2 $  scattering of particles as shown in figure \ref{fig:scattering}. The Mandelstam variables are given by
\be
s=-(p_1+p_2)^2\ , \qquad t=-(p_1-p_3)^2\ , \qquad u=-(p_1-p_4)^2\ .
\ee
The phase-shift $\delta(s,\vec{b})$ has the interpretation of time-delay only when the incoming state 1 is the same as the outgoing state 3 and the incoming state 2 is the same as the outgoing state 4. In the eikonal limit, both incoming particles are highly boosted such that they are moving almost in the null directions. Specifically, the eikonal limit is defined as $s\gg |t|, m_1^2,m_2^2$ \cite{Levy:1969cr}. The phase-shift is defined as the tree-level scattering amplitude expressed in the impact parameter space $\vec{b}$:
\ba\label{phaseshift}
 \delta (s,\vec{b}) = \frac{1}{2 s} \int \frac{d^{2}\vec{q}}{(2 \pi)^{2}} \; e^{i \vec{q} \cdot \vec{b}} M_{\text{tree}}(s, {\vec{q}\;})\ ,
\ea
where in the eikonal limit $t\approx -\vec{q}^{\ 2}$. At first sight one would expect that only ladder diagrams contribute in the eikonal limit and hence the full eikonal amplitude is given by the exponential of the tree level phase shift. In that case, the phase shift has the interpretation of the Shapiro time-delay experienced by either of the particles \cite{Kabat:1992tb,Camanho:2014apa}. The Gao-Wald criteria of asymptotic causality then requires that the time-delay must be non-negative \cite{Gao:2000ga}. 

However, there is no rigorous proof of the eikonal exponentiation. In fact, it is known that the eikonal exponentiation fails for the exchange of particles with spin $J<2$ \cite{Tiktopoulos:1971hi,Cheng:1987ga,Kabat:1992pz}. An elegant physical argument was presented in \cite{Camanho:2014apa} that circumvents this loophole.  When the incoming state 1 is the same as the outgoing state 3 and the incoming state 2 is the same as the outgoing state 4, an eikonal scattering in the regime $s\gg 1/b^2$ can be thought of as a signal transmission problem.\footnote{There is an additional subtlety when particles are metastable. On physical grounds one expects that finite lifetime does not affect the argument because even unstable particles can travel arbitrarily large distances when they are sufficiently boosted. In appendix \ref{appendix:decay} we demonstrate that this expectation is indeed true.}  The signal model then implies that (i) $\delta$ cannot grow faster than $s$, (ii) when $\delta$ grows with $s$ it must be non-negative. Recently, this causality conditions have been used extensively to constrain  interactions of spinning particles \cite{ Camanho:2014apa,Hinterbichler:2017qcl,Bonifacio:2017nnt,Camanho:2016opx,Edelstein:2016nml,Afkhami-Jeddi:2018apj,Chowdhury:2018nfv,Kaplan:2019soo}. 

There is another physical scenario that provides a more direct relation between the tree-level $\delta$ and time-delay by studying  propagation of the particle $1$ in a background with multiple independent shockwaves, each of which is   created by a particle 2. We discuss this set-up in appendix \ref{appendix:decay}.

In the next section, we will argue that metastable HS particles can be coupled to gravity while preserving causality if and only if there is a tower of HS particles in the gravitational sector with masses much below the Planck scale. 

There are other more immediate implications of the above causality condition. We discuss one example that will be useful later. Consider the $2\rightarrow 2 $ scattering of figure \ref{fig:scattering} where all particles are scalars with masses $m_{1,\cdots,4}$. We further restrict to the case where the incoming state 1 is the same as the outgoing state 3 and the incoming state 2 is the same as the outgoing state 4. We fix $t$, and take the limit $|s|\gg |t|, m_i^2$. This is the famous Regge limit in which the amplitude can be parametrized by\footnote{For a review of the Regge limit see \cite{Collins:1977jy,Forshaw:1997dc,Gribov:2003nw,White:2019ggo}. } 
\be\label{eq:regge}
\lim_{|s|\gg |t|, m_i^2} A(s,t) =F(t) (-s)^{j(t)}\ , \qquad \mbox{arg}[s]\neq 0\ ,
\ee
where $j(t)$ is known as the leading Regge trajectory. The amplitude has poles along the positive $s$ axis whenever $s$ hits resonances. The condition $\mbox{arg}[s]\neq 0$ is there to remind ourselves that the large $s$ limit should be taken by avoiding these poles. The regime $t<0$ corresponds to physical high energy small angle scattering. On the other hand, $t>0$ corresponds to unphysical scattering, however, $j(t)$ for positive $t$ contains important information about the spectrum. In particular, solutions of the equation
\be
j(t_J)=J
\ee
for non-negative integer $J$ correspond to particles in the spectrum with mass $m(J)^2=t_J$ and spin $J$. Hence, $F(t)$ has simple poles at $t=t_J$ which enables us to parametrize 
\be
F(t)=\frac{f(t)}{\sin\(\pi j(t)\)}\ .
\ee
Functions $f(t)$ and $j(t)$ have information about the theory. In general, unitarity does impose some constraints such as $f(t)\ge 0$ for $t\ge 0$ and  $j'(t_J) >0$ \cite{Caron-Huot:2016icg}. Given the leading Regge trajectory, one can perform the integral (\ref{phaseshift}) to compute the phase-shift. Let us consider a specific regime: $s \gg 1/b^2, m_i^2$ and $b \gg \log(s)$ (in some appropriate unit). In this regime, the leading contribution to the phase-shift comes from the lightest particle on the leading Regge trajectory
\be
\delta(s,b) \sim f(t=m_0^2)s^{j_0}K_0\(m_0 b\)\ ,
\ee  
where $K_0$ is the Bessel K-function. In addition, $m_0^2=\mbox{min}(t_J)$ is the mass-square of the lightest particles on the Regge trajectory with spin $j_0\equiv j(m_0^2)$. Causality immediately implies
\be\label{eq:causal}
j_0 \le 2\ 
\ee
and $f(t=m_0^2)>0$. Note that the second condition is consistent with unitarity as well.

\subsection{S-Matrix Consistency Conditions}\label{sub:SMatrix}

By studying gravitational scattering of HS particles, we will argue that causality requires that the gravitational sector must contain HS states as well. Moreover, it is known that theories of weakly interacting HS particles are strongly constrained by S-matrix consistency conditions. We will follow \cite{Caron-Huot:2016icg} to explore the asymptotic structure of gravitational scattering amplitudes when the gravity sector contains massive HS states. We make the following assumptions about the S-matrix.\footnote{See \cite{Caron-Huot:2016icg} for a detailed discussion and appendix \ref{appendix:cksz} for a summary.}
\begin{itemize}
\item {Weak coupling -- the Scattering amplitude  $A(s,t)$ is a meromorphic function with simple poles which are located only at resonances.
}
\item{Unitarity-- Unitarity requires that residues admit positive expansions in terms of Legendre polynomials. }
\item{Crossing symmetry--  $A(s,t)=A(t,s)$.}
\item{Regge behavior -- The Regge behavior (\ref{eq:regge}) holds even for $|s|, |t|\rightarrow \infty$ as long as $|t|\ll |s|$.
}
 \item{No accumulation point -- There are finite number of states in the spectrum with masses below any finite mass scale. }
\end{itemize}
The first three assumptions about the S-matrix are self-evident and do not require much explanation. The assumption about the Regge behavior is also reasonable since at high energies all intermediate scales are expected to decouple. Alternatively, one can think of this condition as a definition of the high energy $|t|,|s|\rightarrow\infty$ limit. Finally, the no accumulation point requirement is basically the statement that the theory has a sensible thermodynamic limit. This necessarily implies that  there are finite number of states below any finite energy scale.

In section  \ref{sec:CKSZ} we will show that these S-matrix consistency conditions along with IR causality completely fix the asymptotic spectrum of the gravity sector in HS theories coupled to gravity.

\section{Bounds from Causality}
\label{sec:MainScatteringArgument}

Now let us study flat space scattering in the eikonal limit, with the goal of obtaining an upper bound on the scale $\Lambda_{\rm gr}$ where new higher-spin particles must contribute in the gravity sector.  

We will follow the outline discussed in the introductory section \ref{sec:Introduction}.  We begin by reviewing the fact that there is a unique coupling for a particles $G_J$ of spin $J \geq 3$  and mass $m_J$ with gravitons that remains causal above the scale of their mass.  Through this discussion we will emphasize that we need to study gravitational eikonal scattering at impact parameters in the range $1/m_J \gtrsim b \gtrsim 1/\Lambda_{\rm gr}$.  Then in section \ref{sec:BoundMixing} we establish an important bound on the coupling of distinct  HS particles to  gravity, \ie the $G_J G_{J'}' h$ vertex.  This will then be a crucial ingredient for the coherent state scattering argument in section \ref{sec:ScatteringArgument}.  Finally we conclude in section \ref{sec:ScatteringArgumentResult}.  We detail the kinematics in appendix \ref{app:ScatteringDetails}.

We have defined theories of HS metastable particles in the introduction. Before we proceed, let us comment on two classes of consistent theories with such metastable HS particles.

\begin{enumerate}
\item[A:]  Theories with finite number of HS particles which is weakly coupled, unitary, and causal up to the energy scale $\Lambda_{\rm QFT} \gg m_J$. Causality requires that exchanges of HS particles in any $2\rightarrow 2$ scattering  are suppressed below the cut-off scale \cite{Camanho:2014apa}. Hence, interactions $\langle G_J G_J G_{J'}\rangle$  must be suppressed by appropriate powers of $1/\Lambda_{\rm QFT}$ for $J'\ge 3$ and all $J$. Of course, $G_J$ particles can still interact by exchanging particles with spin $J\le 2$, since these interactions in general are not suppressed by $\Lambda_{\rm QFT}$. It should be noted that it is completely consistent to set $\Lambda_{\rm QCD}=\infty$ and $\langle G_J G_J G_{J'}\rangle=0$ for $J'\ge 3$ and all $J$. For example, massive free HS particles belong in this class.

\item[B:]  Theories with an  infinite number of HS particles with unbounded spin.  Any thermodynamically healthy theory should not have an accumulation point in the spectrum and hence (we will presume that) this scenario necessarily requires $\Lambda_{\rm QFT}=\infty$. These theories are weakly coupled and UV complete. In this case, an infinite tower of HS particles can be exchanged in a $2\rightarrow 2$ scattering such as figure \ref{intro:int} without violating causality. Of course, interactions $\langle G_J G_J G_{J'}\rangle$ are still small because of weak coupling, however, they are not required to be parametrically suppressed by some energy scale. Moreover, if a single $\langle G_J G_J G_{J'}\rangle \neq 0$ for $J'\ge 3$, the CKSZ theorem implies that such a theory must have strings. 
\end{enumerate}

\subsection{Vertex for Graviton Interactions with Higher-Spin Particles}
\label{sec:UniqueJJhVertex}

Consider a tree level scattering process where a spin $J \geq 3$ particle of mass $m_J$ couples to other particles through the exchange of a graviton, as shown in the diagram on the left of figure \ref{fig:BasicDiagrams}.\footnote{Note that only t-channel poles contribute to the phase-shift. Clearly, our set-up is reliable even when $s\gg \Lambda_{\rm QFT}^2$. } This set-up alone, as shown in \cite{Afkhami-Jeddi:2018apj},  is strongly constrained by causality when the exchanged energy is large compared to $m_J$.  In $D > 4$ dimensions this process by itself necessarily violates causality for HS elementary particles.

But in $3+1$ dimensions this process  remains consistent with causality even when the exchanged energy is large compared to $m_J$  provided the on-shell three-point amplitude $\langle G_J G_J h_{\mu\nu}\rangle$ is non-minimal and completely fixed \cite{Afkhami-Jeddi:2018apj}.  Contracting the symmetric tensor polarization indices with null polarization vectors $z_i^\mu$, we can write the amplitude as
\begin{align}\label{bound1}
\langle G_J(p_1,z_1)G_J(p_3,z_3)h(q,z)\rangle_{\rm causal} &=A^2 \sum_{i=1}^{J+1}a_i (z_1\cdot z_3)^{J-i+1}(z_1\cdot q)^{i-1}(z_3\cdot q)^{i-1}\\
+AB &\sum_{i=1}^{J}a_{J+i+1} (z_1\cdot z_3)^{J-i}(z_1\cdot q)^{i-1}(z_3\cdot q)^{i-1}\nonumber
\end{align}
where, $A=(z\cdot p_3)$, $B=(z\cdot z_3)(z_1\cdot q)-(z\cdot z_1)(z_3\cdot q)$ and 
\begin{align}\label{bounds4}
&\frac{a_{n+1}}{a_n}=\frac{(n-J)(n+J-1)}{n(2n-1)}\frac{1}{m_J^2}\ , \qquad n=1,\cdots,J\ , \nonumber\\
&\frac{a_{J+n+2}}{a_{J+n+1}}=\frac{n^2-J^2}{n(2n+1)}\frac{1}{m_J^2}\ , \qquad n=1,\cdots,J-1\ ,
\end{align}
with $a_{J+2}=J a_1$.\footnote{Interactions with general polarization tensors can be obtained from (\ref{bound1}) by acting with the usual Thomas-Todorov operator 
\be
D^{(z)}_\mu= \(1+ z\cdot \p\)\p_\mu-\frac{1}{2}z_\mu \p^2\ ,
\ee
where derivatives are taken with respect to $z^\mu$. 
} 
 Furthermore, consistency of soft limits requires that $a_1=\sqrt{32\pi G_N}=\frac{2}{M_{\rm pl}}$. It was shown in \cite{Kaplan:2019soo} that the same argument, under some additional assumptions, applies to strongly bound composite particles such as glueballs and mesons in large $N$ confining gauge theories. 
 
By considering interference between gravitons and HS particles, it was also shown in \cite{Afkhami-Jeddi:2018apj} that the unique coupling (\ref{bound1}) still violates causality for $J>2$ when there are finite number of HS particles (scenario A with $\Lambda_{\rm QFT}=\infty$). On the other hand, the interference argument is more subtle when there are infinitely many HS particles. We will argue in section \ref{sec:ScatteringArgument} that the interference argument imposes a general bound on $\Lambda_{\rm gr}$. 

Note that in \cite{Arkani-Hamed:2019ymq} a  minimal coupling was discussed, which matches the coupling of gravitons to Kerr black holes with large angular momentum, and arises universally in the classical limit $\hbar \to 0$ with $\hbar J$ fixed.  This coupling matches exactly with the unique interaction (\ref{bound1}).  However, it only describes graviton exchange at scales less than $m_J$, whereas equation (\ref{bound1}) has been engineered to produce causal scattering at energies above $m_J$ for finite $J$.

\subsubsection*{Corrections from Higher Spin exchanges}
The universal coupling was derived by studying the scattering process where a single graviton has been exchanged.  Let us now discuss possible corrections to the preceding  result when we allow other particles in the gravity sector:

\begin{itemize}
\item{Exchange of lower spin ($J<2$) particles: It is not clear if the eikonal exponentiation applies to the exchange of particles with spin $J<2$. However, in the eikonal limit these exchanges are always subleading compared to the graviton exchange and hence can be ignored.}

\item{Exchange of massive spin-2 particles: Massive spin-2 particles, if present, do contribute to the phase shift at the same order as the graviton exchange. However, we can always replace external gravitons by coherent states of gravitons with large occupation number. In that case, as explained in \cite{Camanho:2014apa}, massive spin-2 exchanges do not contribute to the phase-shift for specific polarizations of external gravitons. Moreover, it is also known that the exchange of only massive spin-2 particles (along with the graviton) leads to additional causality violation unless they are accompanied by an infinite tower of finely tuned HS particles. }

\item{Exchange of HS particles ($J>2$): Lorentz invariance of the S-matrix  dictates that massless particles cannot have spin more than two in flat spacetime\cite{Weinberg:1964ew,Weinberg:1980kq,Porrati:2008rm}. On the other hand, massive HS particles are still allowed and they contribute significantly to the phase shift. Of course, as argued in \cite{Camanho:2014apa}, exchange of any finite number of massive HS particles leads to causality violation. However, it is still possible to have a scenario in which an infinite tower of HS particles with finely tuned masses and coupling constants are exchanged without violating causality.}
\end{itemize}

Let us consider the exchange of the lightest massive HS ($J\ge 3$) particle which has mass $m=\Lambda_{\rm gr}$ and spin $J$. In the eikonal limit, exchange of a particle with spin $J$ contributes to the phase shift 
 \be
 \delta \sim s^{J-1} f_J\(\frac{\vec{\p}_{b}}{\Lambda_{\rm gr}}\)K_0\(b \Lambda_{\rm gr}\)\ ,
 \ee
where $K_0$ is the Bessel-K function and the differential operator $f_J\(\frac{\vec{\p}_{b}}{\Lambda_{\rm gr}}\)$ is completely fixed by Lorentz invariance up to some coupling constants. Therefore, this exchange becomes important in the  limit $b\rightarrow 0$. But we can safely ignore this exchange when $b\gtrsim 1/\Lambda_{\rm gr}$.\footnote{To be precise, we can ignore higher spin exchanges for $b \Lambda_{\rm gr} \gtrsim \ln (s/\Lambda)$ where $\Lambda$ is some energy scale set by interactions of the higher spin exchange. We can always take the eikonal limit $s\gg 1/b^2$ without making  $\ln (s/\Lambda)$ much different from order 1. } 

So, in order to derive causality constraints from eikonal scattering thought experiment we should always be in the regime 
\be
\frac{1}{m_J}\gtrsim b \gtrsim \frac{1}{\Lambda_{\rm gr}}\ ,
\ee 
where $\Lambda_{\rm gr}$ is the mass of the lightest massive particle in the gravity sector with spin three or more. Clearly, the strongest constraints can be obtained by setting $b\sim 1/\Lambda_{\rm gr}$ which implies that bounds (\ref{bounds4}) can have corrections which are suppressed by negative powers of $\Lambda_{\rm gr}$.

\subsection{Bounding Graviton-Induced Mixing }
\label{sec:BoundMixing}

Now we derive an important bound on the gravitational coupling to distinct HS particles $G_J$ and $G_{J'}'$, \ie the vertex $G_J G_{J'}' h$.  Our bound is similar to the one derived in \cite{Kaplan:2019soo}. Here we provide a more direct derivation of the same bound that applies to all approximately elementary HS particles in $(3+1)$-dimensions by examining the soft limit of the graviton.

There are two ingredients required to obtain this result.  First we note that as a consequence of soft theorems, this vertex must vanish at low energy, which means that it must grow as a polynomial in momenta at larger energies.  Then we consider scattering of a state which is a linear combination $\alpha G_J + \beta G'_{J'}$, and use causality to bound the phase shift from mixing.  This bound will then require the $G_J G_{J'}' h$ vertex to be suppressed by powers of $\Lambda_{\rm gr}$.

\begin{figure}
\begin{center}
\usetikzlibrary{decorations.markings}    
\usetikzlibrary{decorations.markings}    
\begin{tikzpicture}[baseline=-3pt,scale=0.5]
\begin{scope}[very thick,shift={(4,0)}]
\draw[thin,-latex]  (-3,0) -- (-3.2,2.0);
\draw[thin]  (-3.15,1.5) -- (-3.4,4.0);
\draw[thin]  (-3.2,-2.0) -- (-3,0);
\draw[thin, -latex]  (-3.4,-4)--(-3.15,-1.5) ;
\draw[thin, -latex]  (3,0) -- (3.2,2.0);
\draw[thin]  (3.15,1.5) -- (3.4,4.0);
\draw[thin]  (3.2,-2.0) -- (3,0);
\draw[thin, -latex]  (3.4,-4)--(3.15,-1.5) ;
\draw(-4,1.5)node[above]{$p_3$};
\draw(-4,-1.5)node[below]{$p_1$};
\draw(4,1.5)node[above]{$p_4$};
\draw(4,-1.5)node[below]{$p_2$};
\draw(0,0.3)node[above]{ $q$};
\draw[thin, -latex]  (-0.5,0.4) -- (0.5,0.4);
\draw(0,-0.3)node[below]{ $h_{\mu\nu}$};
\draw(-3.4,-4)node[below]{ $G_J$};
\draw(-3.4,4)node[above]{ $G'_{J'}$};
\draw(3.4,-4)node[below]{ $\psi$};
\draw(3.4,4)node[above]{ $\psi^\dagger$};
\draw(-3.5,0)node[left]{ $\Gamma_{GG'h}$};
\draw(3.5,0)node[right]{ $\Gamma_{\psi\psi^\dagger h}$};
\draw[fill=black](-3,0) circle (0.4);
\draw[fill=black](3,0) circle (0.4);
\draw [domain=-3:3, samples=500] plot (\x, {0.2* cos(3*pi*\x r)});
\end{scope}
\end{tikzpicture}
\end{center}
\caption{\label{fig_mix} \small Eikonal scattering set-up for the phase-shift $\delta_{GG'}$. }
\end{figure}
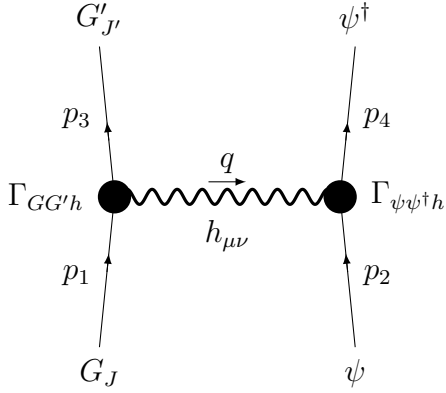

Consider an eikonal scattering between a spectating scalar $\psi$ and an incoming $G_J$ and an outgoing $G'_{J'}$ as shown in figure \ref{fig_mix} (see (\ref{kin4}) for the details of the eikonal kinematics).\footnote{Note that the scalar $\psi$ is not necessary for this argument. We can easily replace $\psi$ by the graviton and make an identical argument. Here, we have introduced the scalar $\psi$ mainly because of two reasons. First, it simplifies the presentation of this section. Secondly, a spectating scalar like $\psi$ can be used as a tool to examine the gravity sector by studying gravitational scattering of  $\psi$ particles. We will do exactly that in section \ref{sec:CKSZ}.} The tree level amplitude consists of the products of three-point functions
\be\label{eq:factor}
M_{\text{tree}}( s,  \vec{q} ) = \frac{\Gamma_{GG'h}(\vec{q})  \Gamma_{\psi \psi^\dagger h}( \vec{q})}{q^2 }\ ,
\ee
where we are assuming that $\psi$ can only interact gravitationally with HS particles. In the above expression, $\Gamma_{GG'h}$ and $\Gamma_{\psi \psi^\dagger h}$ are three-point amplitudes which are in general functions of momenta and polarization tensors.
Using the tree-level amplitude (\ref{eq:factor}), we can compute the phase-shift (\ref{phaseshift}) obtaining 
\begin{align}\label{positive}
\delta_{GG'}& = \frac{1}{2 s} \sum_I\int \frac{d^{2}\vec{q}}{(2 \pi)^{2}} \; e^{i \vec{q} \cdot \vec{b}} \frac{\Gamma_{GG'h}(\vec{q})  \Gamma_{\psi \psi^\dagger h}( \vec{q})}{q^2 } \nonumber\\
&= \frac{1}{4 \pi s} \Gamma_{GG'h}(- i \vec{\partial_b}) \Gamma_{\psi \psi^\dagger h}(- i \vec{\partial_b}) \ln\( \frac{L}{ b}\)\ ,
\end{align}
where $L$ is the IR cut-off. Note that now we can take the exchanged graviton to be on-shell since $\vec{\partial}_b^2$ annihilates  $\ln\( \frac{L}{ b}\)$. This is why we can restrict $\Gamma_{GG'h}$ and $\Gamma_{\psi \psi^\dagger h}$ to be on-shell three-point amplitudes. 

The on-shell three-point function $\Gamma_{\psi \psi^\dagger h}$ is completely fixed by Lorentz invariance and the soft theorem 
\be
\Gamma_{\psi \psi^\dagger h}=\frac{2}{M_{\rm pl}}\epsilon_{\mu\nu}p_2^\mu p_2^\nu
\ee
where $\epsilon_{\mu\nu}$ is the polarization of the graviton. On the other hand, $\Gamma_{GG'h}$ is fixed by Lorentz invariance only up to some coupling constants which we parametrize as a series expansion in $q$ 
\be\label{expansion}
\Gamma_{GG'h}= \frac{2}{M_{\rm pl}}\(\Gamma^{(0)}+\Gamma^{(1)}_\mu q^\mu+\Gamma^{(2)}_{\mu\nu}q^\mu q^\nu+\cdots \)\ .
\ee  
Note that $\Gamma_{GG'h}$ is the on-shell amplitude and hence the above polynomial in $q$ has finite number of terms \cite{Kaplan:2019soo}. Furthermore, $\Gamma^{(0)}=0$ when $G$ and $G'$ are different particles  as a consequence of the soft theorem \cite{Laddha:2017ygw}.\footnote{We are assuming that kinetic mixings between different $G_J$'s are small. See appendix \ref{appendix:soft} for a detailed explanation.} Hence, in the limit $\sqrt{s}\gg \Lambda_{\rm gr}\gg 1/b\gg m_G, m_{G'},m_\psi$, from equation (\ref{positive}) we see that the phase-shift $\delta_{GG'}$ grows with increasing $s$ and $1/b$ at least as fast as 
\be
\delta_{GG'} \sim \frac{s}{M_{\rm pl}^2} \frac{1}{b}\ ,
\ee
We will now argue that the $b$-dependence of this phase shift is in tension with causality.

\begin{figure}
\begin{center}
\usetikzlibrary{decorations.markings}    
\usetikzlibrary{decorations.markings}    
\begin{tikzpicture}[baseline=-3pt,scale=0.45]
\begin{scope}[very thick,shift={(4,0)}]
\draw[thin,-latex]  (-3,0) -- (-3.2,2.0);
\draw[thin]  (-3.15,1.5) -- (-3.4,4.0);
\draw[thin]  (-3.2,-2.0) -- (-3,0);
\draw[thin, -latex]  (-3.4,-4)--(-3.15,-1.5) ;
\draw[thin, -latex]  (3,0) -- (3.2,2.0);
\draw[thin]  (3.15,1.5) -- (3.4,4.0);
\draw[thin]  (3.2,-2.0) -- (3,0);
\draw[thin, -latex]  (3.4,-4)--(3.15,-1.5) ;
\draw(-4,1.5)node[above]{$p_3$};
\draw(-4,-1.5)node[below]{$p_1$};
\draw(4,1.5)node[above]{$p_4$};
\draw(4,-1.5)node[below]{$p_2$};
\draw(0,0.3)node[above]{ $q$};
\draw(0,-0.3)node[below]{ $h_{\mu\nu}$};
\draw(-3.4,-4)node[below]{ $\alpha G_J+ \beta G'_{J'}$};
\draw(-3.4,4)node[above]{ $\alpha' G_J+ \beta' G'_{J'}$};
\draw(3.4,-4)node[below]{ $\psi$};
\draw(3.4,4)node[above]{ $\psi^\dagger$};
\draw[fill=black](-3,0) circle (0.4);
\draw[fill=black](3,0) circle (0.4);
\draw [domain=-3:3, samples=500] plot (\x, {0.2* cos(3*pi*\x r)});
\draw[thin, -latex]  (-0.5,0.4) -- (0.5,0.4);
\end{scope}
\end{tikzpicture}
\end{center}
\caption{\label{fig_mixG} \small A setup to bound the graviton induced mixing. The leading non-trivial contribution still comes from a graviton exchange.}
\end{figure}
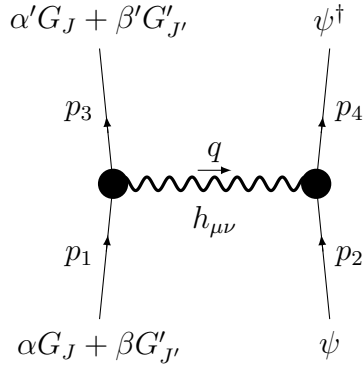

Consider an eikonal scattering: $1,2\rightarrow 3,4$, where, 1 and 3 are linear combinations $\alpha G_J+ \beta G'_{J'}$ and $\alpha' G_J+ \beta' G'_{J'}$ respectively with real coefficients $\alpha,\alpha',\beta,\beta'$ (see figure \ref{fig_mixG}). Particles 2 and 4 are either a spectating scalar $\psi$ or a graviton $h_{\mu\nu}$. Causality now  can be expressed as semi-definiteness of the phase shift matrix $\delta_{13}$:
\begin{align}\label{matrix}
\delta_{13}\equiv \left(
\begin{array}{cc}
\delta_{GG} & \delta_{GG'}\\ 
\delta_{GG'}^* & \delta_{G'G'}
\end{array} 
\right)\succeq 0\ .
\end{align}
This condition imposes  a bound on $\delta_{GG'}$:
\be\label{eq:intf_bound}
|\delta_{GG'}|^2 \le \delta_{GG}\delta_{G'G'}\ .
\ee
The causality conditions (\ref{bound1}) and (\ref{bounds4}) imply that for all external polarizations
\be
\delta_{GG}= \frac{4s}{M_{\rm pl}^2} \ln \left(\frac{L}{b} \right)\ , \qquad \delta_{G'G'}=  \frac{4s}{M_{\rm pl}^2} \ln \left(\frac{L}{b} \right)\ 
\ee
up to some overall factors that depend on polarizations \cite{Afkhami-Jeddi:2018apj}. Hence, $\delta_{GG'}$  cannot grow faster than $\frac{s}{M_{\rm pl}^2} \ln \left(\frac{L}{b} \right)$ in the limit $\sqrt{s}\gg \Lambda_{\rm gr}\gg 1/b\gg m_G, m_{G'},m_\psi$.

This is clearly inconsistent with (\ref{eq:intf_bound}) unless $\Gamma^{(i)}$s are suppressed by $\Lambda_{\rm gr}$ which is the scale at which new HS states show up in the gravity sector. We can set $b\sim 1/\Lambda_{\rm gr}$, where $\Lambda_{\rm gr}\gg m_G, m_{G'}, m_\psi$. In this limit, the bound (\ref{eq:intf_bound}) 
implies that the on-shell three-point amplitude 
\be\label{bound2}
|\Gamma_{GG'h}| \lesssim \frac{1}{M_{\rm pl}} \frac{\ln(\Lambda_{\rm gr} L)}{\Lambda_{\rm gr}^{n}} \qquad \text{with} \qquad n\ge 1 \ .
\ee
The $G G' h$ vertex must be suppressed by powers of the $\Lambda_{\rm gr}$ scale, so the  amplitudes that change the identity of $G$ must be suppressed compared to amplitudes that preserve $G$.

These bounds are interesting in themselves, but  they do not put any upper bound on $\Lambda_{\rm gr}$, and would be consistent were we to simply eliminate all mixing amplitudes.  To demonstrate an upper bound on $\Lambda_{\rm gr}$, we will need to combine these results with the analysis of a different kind of amplitude.

\subsection{Scattering Argument Using Coherent States}
\label{sec:ScatteringArgument}

Now we will study a scattering experiment involving coherent states formed from superpositions of $G_J$ and gravitons $h$.  Let us first identify an inequality on the phase shifts associated with this process, and then we will compute them.

\begin{figure}
\begin{center}
\usetikzlibrary{decorations.markings}    
\usetikzlibrary{decorations.markings}    
\begin{tikzpicture}[baseline=-3pt,scale=0.45]
\begin{scope}[very thick,shift={(4,0)}]
\draw[thin,-latex]  (-3,0) -- (-3.2,2.0);
\draw[thin]  (-3.15,1.5) -- (-3.4,4.0);
\draw[thin]  (-3.2,-2.0) -- (-3,0);
\draw[thin, -latex]  (-3.4,-4)--(-3.15,-1.5) ;
\draw[thin, -latex]  (3,0) -- (3.2,2.0);
\draw[thin]  (3.15,1.5) -- (3.4,4.0);
\draw[thin]  (3.2,-2.0) -- (3,0);
\draw[thin, -latex]  (3.4,-4)--(3.15,-1.5) ;
\draw[very thick, -latex]  (-3,0)--(0,0) ;
\draw[very thick ]  (-0.1,0)--(3,0) ;
\draw(-2.5,1.5)node[above]{$p_3$};
\draw(-2.5,-1.5)node[below]{$p_1$};
\draw(2.5,1.5)node[above]{$p_4$};
\draw(2.5,-1.5)node[below]{$p_2$};
\draw(0,0.2)node[above]{ $q$};
\draw(0,-0.3)node[below]{ $I$};
\draw(-3.4,-4)node[below]{ $1$};
\draw(-3.4,4)node[above]{ $3$};
\draw(3.4,-4)node[below]{ $2$};
\draw(3.4,4)node[above]{ $4$};
\draw(-10,0)node[right]{ $\sum_{I}$};
\draw(3.4,3.5)node[right]{ $h+G_J$};
\draw(3.4,-3.5)node[right]{ $h+G_J$};
\draw(-3.4,3.5)node[left]{ $\alpha h +\beta G_J$};
\draw(-3.4,-3.5)node[left]{ $\alpha' h +\beta' G_J$};
\end{scope}
\end{tikzpicture}
\end{center}
\caption{\label{fig_int} \small Graviton interference bound: in-states are linear combinations of $G_J$ with $J\ge 3$ and the graviton $h$. }
\end{figure}
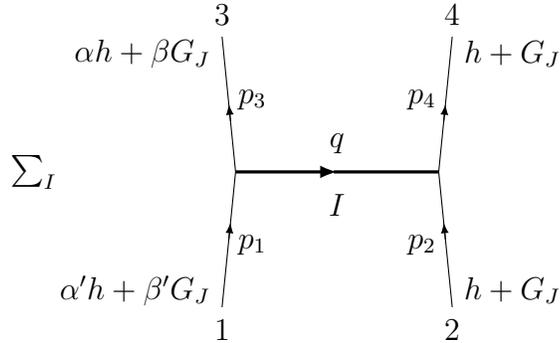

Consider the eikonal scattering as shown in figure \ref{fig_int} -- states 1 and 3 are linear combinations of $G_J$ and the graviton: $\alpha h +\beta G_J$ and $\alpha' h+ \beta' G_J$ respectively, where $\alpha,\alpha',\beta,\beta'$ are some arbitrary real coefficients. States 2 and 4 are a fixed combination of $G_J$ and the graviton: $ h + G_J$. Positivity of the phase-shift, as shown in \cite{Afkhami-Jeddi:2018apj}, can now be expressed as an interference bound. Here we are including graviton induced mixings, since $G_J$'s are only approximately elementary. Hence, the interference bound becomes
\be\label{intf_bound}
|\delta_{Gh}^{hG}+\delta_{GG}^{hh}+\delta_{GG}^{hG}|^2 \le \(\delta_{hh}^{hh}+\delta_{hG}^{hG}\)\(\delta_{GG}^{GG}+\delta_{Gh}^{Gh}+\delta_{GG}^{Gh}+\delta_{Gh}^{GG}\)\ ,
\ee
where $\delta_{12}^{34}$ represents the phase shift for the process $12\rightarrow 34$. Note that we ignored $\delta_{hh}^{hG}$ and similar terms since they are suppressed by $\frac{1}{M_{\rm pl}^3}$. The above expression can be manipulated into a slightly simpler inequality\footnote{Note that phase-shifts such as $\delta_{GG}^{hG}$ also obey interference bounds individually $|\delta_{GG}^{hG}|^2\le \delta_{GG}^{hh}\delta_{GG}^{GG}$. }
\be\label{qcd_bound}
 |\delta_{Gh}^{hG}+\delta_{GG}^{hh}|\le \sqrt{\delta_{hh}^{hh}+\delta_{hG}^{hG}}\(\sqrt{\delta_{GG}^{GG}}+\sqrt{\delta_{Gh}^{Gh}}\)+\sqrt{\delta_{GG}^{GG}\delta_{hG}^{hG}}\ .
\ee
This imposes a bound on the diagram on the right of figure \ref{fig:CausalityConstrainedScatteringDiagrams}. Note that  in the eikonal limit $\delta_{Gh}^{hG} \neq \delta_{hG}^{hG}$, because the process $12 \to 34$ and $12 \to 43$ have very different kinematics when $s \gg t$.  

Before we utilize the bound (\ref{qcd_bound}) to derive constraints, let us make a comment about the  diagram on the right of figure \ref{fig:CausalityConstrainedScatteringDiagrams}.  This diagram is related to certain dispersion relations known as superconvergent sum rules \cite{DEALFARO1966576,PhysRev.165.1803,Kologlu:2019bco}. Consider propagation of a $G_J$ particle through multiple gravitational shockwaves, similar to the scenario of figure \ref{nshock}. In \cite{Kologlu:2019bco}, authors proposed that  in any UV complete theory of gravity coincident gravitational shocks commute. Commutativity of coincident shocks can be alternatively stated as vanishing of certain superconvergent sum rules in gravity. This condition strongly constraints the  diagram on the right of figure \ref{fig:CausalityConstrainedScatteringDiagrams}. Theories of HS particles, when coupled to gravity, are completely consistent with the shock commutativity condition below the energy scale $\Lambda_{\rm gr}$ for all spins, provided  gravitational interactions obey (\ref{bounds4}) and (\ref{bound2}).

We now compute the various phase-shifts that appear in the inequality above. We detail the kinematics in appendix \ref{app:ScatteringDetails}. First note that contact diagrams or exchanges in other channels do not contribute to eikonal phase-shifts at the tree-level for $\vec{b}\neq 0$. This implies we only need to consider diagrams that contain on-shell particles in the $t$-channel. This simplifies computations greatly.

The main point will be that in the eikonal limit $\delta_{GG}^{hh}$ and $\delta_{Gh}^{hG}$ on the LHS of equation (\ref{qcd_bound}) grow quickly with increasing energy, as a consequence of the diagram at right in figure \ref{fig:CausalityConstrainedScatteringDiagrams}.  In contrast, the phase shifts on the RHS only grow in proportion to $s$.  This eventually violates the inequality, putting an upper bound on $\Lambda_{\rm gr}$.

All-graviton vertices are fixed by causality, Lorentz invariance, and soft theorems. The dominant contribution to $\delta_{hh}^{hh}$ in the eikonal limit comes from a single graviton exchange\footnote{We should note that phase-shifts are in general functions of polarizations of external particles. However, we fix external polarization as given in \ref{app:ScatteringDetails} from now on and ignore this functional dependence. } 
\be
\delta_{hh}^{hh}=\frac{4s}{M_{\rm pl}^2} \ln \left(\frac{L}{b} \right)+\frac{s}{M_{\rm pl}^2}\O\(\frac{1}{\Lambda_{\rm gr}^2}\)\ ,
\ee 
where higher-derivative interactions are suppressed by $1/\Lambda_{\rm gr}$. At the leading order, $G_J$ interacts with gravitons only gravitationally. Moreover, the $G_J G_J h$ interaction is completely fixed   by causality (\ref{bound1}) which implies that in the eikonal limit
\be
\delta_{hG}^{hG}=\delta_{Gh}^{Gh}=\frac{4s}{M_{\rm pl}^2} \ln \left(\frac{L}{b} \right)+\frac{s}{M_{\rm pl}^2}\O\(\frac{1}{\Lambda_{\rm gr}}\)\ .
\ee
Of course, $G_J$ can interact with gravitons non-gravitationally by exchanging a set of $G_J$ particles with different masses and spins. However, these interactions are subleading since they are suppressed by an additional factor of $\lambda  \ll 1$, as defined in equation (\ref{eq:interaction}).

\begin{figure}
\begin{center}
\usetikzlibrary{decorations.markings}    
\usetikzlibrary{decorations.markings}    
\begin{tikzpicture}[baseline=-3pt,scale=0.40]
\begin{scope}[very thick,shift={(4,0)}]
\draw[thin,-latex]  (-3,0) -- (-3.2,2.0);
\draw[thin]  (-3.15,1.5) -- (-3.4,4.0);
\draw[thin]  (-3.2,-2.0) -- (-3,0);
\draw[thin, -latex]  (-3.4,-4)--(-3.15,-1.5) ;
\draw[thin, -latex]  (3,0) -- (3.2,2.0);
\draw[thin]  (3.15,1.5) -- (3.4,4.0);
\draw[thin]  (3.2,-2.0) -- (3,0);
\draw[thin, -latex]  (3.4,-4)--(3.15,-1.5) ;
\draw[very thick, -latex]  (-3,0)--(0,0) ;
\draw[very thick ]  (-0.1,0)--(3,0) ;
\draw(-4,1.5)node[above]{$p_3$};
\draw(-4,-1.5)node[below]{$p_1$};
\draw(4,1.5)node[above]{$p_4$};
\draw(4,-1.5)node[below]{$p_2$};
\draw(0,0.3)node[above]{ $\sum G'_{J'}$};
\draw(5,0)node[right]{ $= g_J^2\(\frac{s}{m_J^2}\)^{a_{J}}$};
\draw(-3.4,-4)node[below]{ $G_J$};
\draw(-3.4,4)node[above]{ $G_J$};
\draw(3.4,-4)node[below]{ $G_J$};
\draw(3.4,4)node[above]{ $G_J$};
\draw[fill=black](-3,0) circle (0.4);
\draw[fill=black](3,0) circle (0.4);
\end{scope}
\end{tikzpicture}
\caption{\small Phase-shifts associated with non-gravitational interaction between HS particles. This parametrization is defined for specific polarizations of external particles that are given in \ref{app:ScatteringDetails}. We also set impact parameter $b=1/m_J$.}\label{p4}
\end{center}
\end{figure}
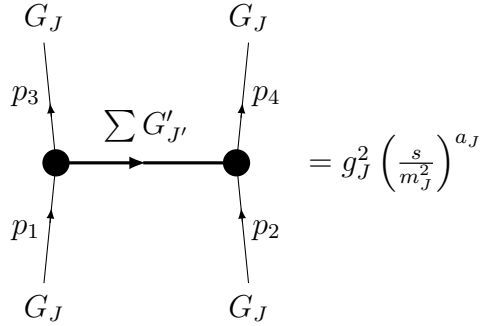

Two different processes contribute to $\delta_{GG}^{GG}$. After imposing the causality constraints (\ref{bound1}),  the eikonal phase shift for external polarizations \ref{app:ScatteringDetails} is given by 
\be\label{GGGG}
\delta_{GG}^{GG}= \frac{4s}{M_{\rm pl}^2} \ln \left(\frac{L}{b} \right)+g_J^2\(\frac{s}{m_J^2}\)^{a_{J}}
\ee
where $a_{J}$ and $g_J$ are dimensionless coefficients.  The first term comes from the graviton exchange and the second term is from non-gravitational interactions, as shown in figure \ref{p4}. We parametrize the phase-shift for non-gravitational interactions in the above way for impact parameter $b=1/m_J$. Coefficients $a_{J}$ and $g_J$ are theory dependent. However, we still know that $|g_J| \ll 1$ because of weak coupling. Moreover, causality requires that $a_{J}\le 1$. 

Let us make few more comments about the non-gravitational part of (\ref{GGGG}). If there are finite number of HS particles in the non-gravitational sector (alternatively, a class A theory with $\Lambda_{\rm QFT}=\infty$), then only particles of spin 0,1, and 2 can contribute in figure \ref{p4}. This follows from the fact that the sector $\{G_J\}$ is a consistent theory even before we couple it to gravity. Hence, the phase-shift associated with the non-gravitational interaction between $G_J$'s must not grow faster than $s$. On the other hand, when the sector $\{G_J\}$ contains an infinite number of HS particles, the process \ref{p4} may get contributions from an infinite tower of HS exchanges fine tuned to be consistent with causality  $a_{J}\le 1$.

In the eikonal limit the dominant contributions to $\delta_{Gh}^{hG}$ and $\delta_{GG}^{hh}$ come from the $G_J$-exchange. Moreover, for large but finite $\Lambda_{\rm gr}$, an infinite tower of HS exchanges can also contribute to the above process (see figure \ref{figure_mix}), however, they must be suppressed by positive powers of $1/\Lambda_{\rm gr}$ because of (\ref{bound2}). In particular, after imposing constraints (\ref{bounds4}), in  the limit $\sqrt{s}\gg m_J, 1/b$ we find that 
\be \label{eq:GGhhGrowth}
\delta_{Gh}^{hG}=\delta_{GG}^{hh}= f_{J}\frac{s^{J-1}}{M_{\rm pl}^2} \frac{e^{-2i(J-2)\theta}}{b^{2(J-2)}m_J^{4(J-2)}} +\frac{s m_J^2}{M_{\rm pl}^2\Lambda_{\rm gr}^2}\sum_{J'} \(\frac{ s}{m_J^2} \)^{J'-2} \tilde{f}_{J'}(\vec{b},m_{J'})
\ee
where $\cos\theta=\hat{b}\cdot \hat{x}$ and $f_{J}$ is an $\O(1)$ numerical coefficient.\footnote{Let us emphasize that $f_{J}$ is completely fixed by (\ref{bounds4}) and hence cannot be tuned.} In the above equation, the first term on the right hand side is exact. On the other hand, the second term is a parametrization of the second process in figure \ref{figure_mix}. Coefficient functions $ \tilde{f}_{J'}(\vec{b},m_{J'})$ are theory dependent and they can be suppressed by additional powers of $1/\Lambda_{\rm gr}$. However, we do not actually need the exact functional forms of $ \tilde{f}_{J'}(\vec{b},m_{J'})$ to derive our final bounds as long as $ \tilde{f}_{J'}(\vec{b},m_{J'})$ are not too large. Of course, only $m_{J'}\sim m_J$ can significantly contribute in the above sum. All exchanges with $m_{J'}\gg m_J$ will be exponentially suppressed $\tilde{f}_{J'}(\vec{b},m_{J'})\sim e^{-m_{J'}/m_J}$ for $b= 1/m_J$. We further assume that the infinite sum of the second term converges for $\sqrt{s}\ll \Lambda_{\rm gr}$ and $m_J\sim m_{J'}\sim 1/b $. 

\begin{figure}[h]
\usetikzlibrary{decorations.markings}    
\usetikzlibrary{decorations.markings}    
\begin{center}

\begin{tikzpicture}[baseline=3pt,scale=0.40]

\begin{scope}[very thick,shift={(4,0)}]
\draw[thin,-latex]  (-3,0) -- (-3.2,2.0);
\draw[thin]  (-3.15,1.5) -- (-3.4,4.0);
\draw[thin]  (-3.2,-2.0) -- (-3,0);
\draw[thin, -latex]  (-3.4,-4)--(-3.15,-1.5) ;
\draw[thin, -latex]  (3,0) -- (3.2,2.0);
\draw[thin]  (3.15,1.5) -- (3.4,4.0);
\draw[thin]  (3.2,-2.0) -- (3,0);
\draw[thin, -latex]  (3.4,-4)--(3.15,-1.5) ;
\draw[very thick, -latex]  (-3,0)--(0,0) ;
\draw[very thick ]  (-0.1,0)--(3,0) ;
\draw(-4,1.5)node[above]{$p_3$};
\draw(-4,-1.5)node[below]{$p_1$};
\draw(4,1.5)node[above]{$p_4$};
\draw(4,-1.5)node[below]{$p_2$};
\draw(0,0.3)node[above]{ $G_{J}$};

\draw(5,0)node[right]{ $= f_{J}\frac{s^{J-1}}{M_{\rm pl}^2} \frac{e^{-2i(J-2)\theta}}{b^{2(J-2)}m_J^{4(J-2)}} $};
\draw(-3.4,-4)node[below]{ $G_J$};
\draw(-3.4,4)node[above]{ $h$};
\draw(3.4,-4)node[below]{ $h/G_J$};
\draw(3.4,4)node[above]{ $G_J/h$};
\draw(-8,0)node[left]{ (a)};
\draw[fill=black](-3,0) circle (0.4);
\draw[fill=black](3,0) circle (0.4);
\end{scope}
\end{tikzpicture}
\end{center}

\begin{tikzpicture}[baseline=3pt,scale=0.40]
\hspace*{0.18\linewidth}
\begin{scope}[very thick, shift={(6,4)}]
\draw[thin,-latex]  (-3,0) -- (-3.2,2.0);
\draw[thin]  (-3.15,1.5) -- (-3.4,4.0);
\draw[thin]  (-3.2,-2.0) -- (-3,0);
\draw[thin, -latex]  (-3.4,-4)--(-3.15,-1.5) ;
\draw[thin, -latex]  (3,0) -- (3.2,2.0);
\draw[thin]  (3.15,1.5) -- (3.4,4.0);
\draw[thin]  (3.2,-2.0) -- (3,0);
\draw[thin, -latex]  (3.4,-4)--(3.15,-1.5) ;
\draw[very thick, -latex]  (-3,0)--(0,0) ;
\draw[very thick ]  (-0.1,0)--(3,0) ;
\draw(-4,1.5)node[above]{$p_3$};
\draw(-4,-1.5)node[below]{$p_1$};
\draw(4,1.5)node[above]{$p_4$};
\draw(4,-1.5)node[below]{$p_2$};
\draw(0,0.3)node[above]{ $\sum G'_{J'}$};
\draw(5,0)node[right]{ $=\frac{s m_J^2}{M_{\rm pl}^2\Lambda_{\rm gr}^2}\sum_{J'} \(\frac{ s}{m_J^2} \)^{J'-2} \tilde{f}_{J'}(\vec{b},m_{J'})$};
\draw(-3.4,-4)node[below]{ $G_J$};
\draw(-3.4,4)node[above]{ $h$};
\draw(3.4,-4)node[below]{ $h/G_J$};
\draw(3.4,4)node[above]{ $G_J/h$};
\draw(-8,0)node[left]{ (b)};
\draw[fill=black](-3,0) circle (0.4);
\draw[fill=black](3,0) circle (0.4);
\end{scope}
\end{tikzpicture}

\caption{\small In the eikonal limit the dominant contribution to $\delta_{Gh}^{hG}$ and $\delta_{GG}^{hh}$ come from the $G_J$-exchange, as shown in (a). For large but finite $\Lambda_{\rm gr}$, an infinite tower of HS exchanges can contribute to $\delta_{Gh}^{hG}$ and $\delta_{GG}^{hh}$ which is shown in (b). Note that the process with a graviton exchange is suppressed by  $1/M_{\rm pl}^3$.}\label{figure_mix}
\end{figure}

At finite $g_J$ the HS particles $G_J$ does decay to other states.  This process will have a parametric rate $\propto g_J^2$, but it includes unknown form factors that are theory dependent.  So we require $|g_J| \ll 1$ to ensure that the scattering process occurs before the particle decays. Moreover, the interference bound requires the incoming state 1 and the outgoing state 3 to be a linear combination of two different particles. In general, these particles can have different masses and hence different momenta. So, if we wait for a long time, two different incoming particles will move away from each other. This implies that we can trust our interference bound only if $s\gg m_J^2,\frac{1}{b^2}$.  For a more extensive discussion of these and other subtleties see appendix \ref{appendix:decay}.

\subsection{Implications}
\label{sec:ScatteringArgumentResult}

\subsubsection*{Non-interacting metastable HS particles}

We first prove a simple theorem:
\begin{quote}
{\it A metastable HS particle of mass $m_J$, when coupled to gravity, is ruled out in the limit $g_J\rightarrow 0$ and $\Lambda_{\rm gr}\gg m_J$.}
\end{quote}
In this limit  we can ignore most of the processes described above if we are in the regime: $\Lambda_{\rm gr}\gg \sqrt{s}\gg m_J$. In particular, for $b= 1/m_J$ we have
\begin{align}
&\delta_{hh}^{hh}=\delta_{hG}^{hG}=\delta_{Gh}^{Gh}\approx\delta_{GG}^{GG}\approx \frac{4s}{M_{\rm pl}^2} \ln \left(\frac{L}{b} \right)\ ,\nonumber\\
&\delta_{Gh}^{hG}=\delta_{GG}^{hh}\approx  f_{J}\frac{s^{J-1}}{M_{\rm pl}^2} \frac{e^{-2i(J-2)\theta}}{b^{2(J-2)}m_J^{4(J-2)}}
\end{align}
which clearly violates the inequality (\ref{qcd_bound}) for any $J\ge 3$. This rules out even a single free massive HS particle if $\Lambda_{\rm gr}\gg m_J$. 
We can restore causality if 
\begin{itemize}
\item $\Lambda_{\rm gr}\gg m_J$ and $g_J$ is small but nonzero, or
\item $g_J\rightarrow 0$  but $\Lambda_{\rm gr}\sim m_J$.
\end{itemize}

\subsubsection*{A bound on $\Lambda_{\rm gr}$ for weak coupling $|g_J|\lesssim \frac{m_J}{M_{\rm pl}}$}
In the latter case, causality necessarily requires the introduction of HS states in the gravity sector at an energy scale comparable to the mass $m_{\rm min}$ of the lightest HS particle in the $\{G_J\}$-sector.  Here it is easy to see why $\{G_J\}$-sector cannot cure their own pathologies, as their interactions are vanishingly weak. Specifically, in the limit $|g_J| \ll \frac{m_J}{M_{\rm pl}}$ and $\Lambda_{\rm gr}\gg \sqrt{s}\gg m_J$, we have 
\begin{align}
&\delta_{hh}^{hh}=\delta_{hG}^{hG}=\delta_{Gh}^{Gh}\approx \delta_{GG}^{GG} \approx \frac{4s}{M_{\rm pl}^2} \ln \left(\frac{L}{b} \right)+\frac{s}{M_{\rm pl}^2}\O\(\frac{1}{\Lambda_{\rm gr}}\)\ ,\\
&\delta_{Gh}^{hG}=\delta_{GG}^{hh}\approx f_{J}\frac{s^{J-1}}{M_{\rm pl}^2} \frac{e^{-2i(J-2)\theta}}{b^{2(J-2)}m_J^{4(J-2)}} +\frac{s m_J^2}{M_{\rm pl}^2\Lambda_{\rm gr}^2}\sum_{J'} \(\frac{ s}{m_J^2} \)^{J'-2} \tilde{f}_{J'}(\vec{b},m_{J'})\nonumber
\ .
\end{align}
The bound (\ref{qcd_bound}) in this limit can only be satisfied for a $G_J$ particle if 
\be\label{eq:weakbound}
\Lambda_{\rm gr} \lesssim m_J
\ee
such that the correction terms are large enough to restore causality. Hence, when $g_J\rightarrow 0$ for all $G_J$, $\Lambda_{\rm gr} \lesssim m_{\rm min}$.

\subsubsection*{A weak gravity condition for HS particles}
 Any HS particle of mass $m_J$ coupled to gravity can have a QFT description only if $\Lambda_{\rm gr}\gg m_J$. From the preceding discussion, it is clear that $\Lambda_{\rm gr}\gg m_J$ requires $g_J$ to be small but finite. In particular, it must obey $|g_J|\gg \frac{m_J}{M_{\rm pl}}$.

What bounds can we obtain on $\Lambda_{\rm gr}$ for finite $g_J$? We will first derive a very conservative bound, which is equivalent to the statement that all HS particle $G_J$'s must have non-gravitational interactions that are stronger than their gravitational interactions.  However, intuitively it's clear that the $s^{J-1}$ growth of equation (\ref{eq:GGhhGrowth}) is very much at odds with the softer dependence on $s$ of the other phase shifts.  This suggests that we should be able to obtain a stronger bound, and we will argue for one below.

In the limit $\Lambda_{\rm gr}\gg \sqrt{s}\gg m_J,\frac{1}{b}$, we can approximate 
\begin{align}\label{ps1}
&\delta_{hh}^{hh}=\delta_{hG}^{hG}=\delta_{Gh}^{Gh}\approx \frac{4s}{M_{\rm pl}^2} \ln \left(\frac{L}{b} \right)\ ,\nonumber\\
&\delta_{GG}^{GG}=\delta_{\rm grav}+\delta_{\rm non-grav}\approx \frac{4s}{M_{\rm pl}^2} \ln \left(\frac{L}{b} \right)+\delta_{\rm non-grav}\ ,\nonumber\\
&\delta_{Gh}^{hG}=\delta_{GG}^{hh}\approx f_{J}\frac{s}{M_{\rm pl}^2}\( \frac{s}{m_J^{2}}\)^{J-2}e^{-2i(J-2)\theta}\ ,
\end{align}
We require $|\delta_{Gh}^{hG}|\ll 1$ so that we are in the weakly coupled regime. This can always be satisfied for some $\sqrt{s}$ within a window between $\Lambda_{\rm gr}$ and $m_J$, provided $J$ is not infinitely large. In the regime $\Lambda_{\rm gr}\gg \sqrt{s}\gg m_J,\frac{1}{b}$,  we see that $ |\delta_{Gh}^{hG}| \gg |\delta_{hh}^{hh}|$ for  $J\ge 3$.  Hence, the bound (\ref{qcd_bound}) necessarily requires 
\be\label{eq:wg}
\delta_{\rm non-grav} \gg \delta_{\rm grav}
\ee
for all $G_J$ particles with $J\ge 3$. This last very conservative inequality essentially just states that the non-gravitational interactions of $G_J$ particles must be stronger than their gravitational interactions.  Moreover, the bound (\ref{qcd_bound}) in this limit can be rewritten approximately as
\be\label{qcd_bound2} 
2 |\delta_{Gh}^{hG}|\le (1+ \sqrt{2})\sqrt{\delta_{GG}^{GG}\delta_{hh}^{hh}} \ .
\ee
Note that for $b= 1/m_J$, in the limit $\Lambda_{\rm gr}\gg \sqrt{s}\gg m_J$, we can parametrize
\be
\delta_{\rm non-grav}=g_J^2\(\frac{s}{m_J^2}\)^{a_{J}}\ .
\ee
The exponent $a_{J} \le 1$ is determined by the Regge behavior of scattering in the $\{G_J\}$-sector.  The fact that the second term in $\delta_{GG}^{GG}$ dominates over the first term strongly effects the character of the constraints. In particular, the bound (\ref{qcd_bound2}) implies the constraint
\be\label{con1}
\frac{1}{|g_J|}\ll  \frac{M_{\rm pl}}{m_{J}} \(\frac{m_J}{\sqrt{s}}\)^{1-a_{J}}
\ee
This can be re-written as a bound on the HS scale by evaluating $\sqrt s \sim \Lambda_{\rm gr}$, so\footnote{For theories with $\Lambda_{\rm QFT}< \Lambda_{\rm gr}$, at first sight one may think that $\sqrt{s}\lesssim \Lambda_{\rm QFT}$. However, this is not exactly true and one can still take $ \Lambda_{\rm gr}\gtrsim \sqrt{s}> \Lambda_{\rm QFT}$ as long as $\sqrt{|t|}\ll  \Lambda_{\rm QFT}$. Of course, the theory may not be described by a convenient effective action, since the putative cutoff will be of order $ \Lambda_{\rm QFT}$.   Nevertheless, the phase-shift computations depend only on t-channel residues and hence they remain reliable even when $\sqrt{s}> \Lambda_{\rm QFT}$.} 
\be\label{s4}
\Lambda_{\rm gr} \lesssim m_J\(\frac{|g_J| M_{\rm pl}}{m_J}\)^{\frac{1}{1-a_{J}}}\ .
\ee
The above bound specifically requires HS particles in the {\it gravity sector} at $\Lambda_{\rm gr}$.

Given a theory of metastable HS particles coupled to gravity, we can obtain a low energy QFT description for a set of light HS particles by integrating out states above $\Lambda_{\rm gr}$. This necessarily requires $|g_J|\gtrsim \frac{m_J}{M_{\rm pl}}$ for all HS particles in the $\{G_J\}$-sector with masses $m_J \ll \Lambda_{\rm gr}$. These particles must also obey the condition (\ref{eq:wg}), as shown in figure \ref{fig:weakgravity}. This establishes the weak gravity condition for metastable HS particles.

\subsubsection*{A stronger bound on $\Lambda_{\rm gr}$ for $|g_J|\gtrsim \frac{m_J}{M_{\rm pl}}$}

Of course, the bound (\ref{s4}) is just a necessary condition, and it is far from sufficient.  Intuitively, it would seem that the $\delta_{GG}^{hh}$ phase shift grows much more quickly with $s$ than the other phase shifts, and this should provide a much stronger bound. From equation (\ref{qcd_bound2}) we obtain a \emph{parametric} bound
\be
\Lambda_{\rm gr} \lesssim m_J \(\frac{|g_J| M_{\rm pl}}{ m_J}\)^{\gamma(J)} 
\label{eq:StrongBound}
\ee
where  
\be\label{gamma}
\gamma(J)= \frac{1}{1-a_{J}+2(J-2)}\le  \frac{1}{2(J-2)}\ .
\ee
The inequality (\ref{eq:StrongBound}), together with (\ref{eq:weakbound}), are bounds that all HS particles in the $\{G_J\}$-sector of both class A and class B theories must obey. This is summarized in figure \ref{fig:intro}. The optimal bound on $\Lambda_{\rm gr}$ is obtained by minimizing the right hand sides of (\ref{eq:StrongBound}) and (\ref{eq:weakbound}). However, the optimal bound depends importantly on unknown, theory dependent form factors that determine the non-gravitational scattering amplitudes. Since  the relevant form-factors are theory dependent,  we cannot determine the optimal bound in general. In any case, the most conservative estimate is obtained for $J=3, a_{J} \approx 1$ which predicts $\gamma \approx  1/2$.

\subsubsection*{Elementary HS particles}

We discuss a special case to emphasize that the bound (\ref{eq:StrongBound}) may still be far from sufficient. We revisit the theorem proved in \cite{Afkhami-Jeddi:2018apj} by considering a theory of finite number of elementary HS particles along with particles with spin $J\le 2$. In the language of this paper, such a theory is simply a class A theory with $\Lambda_{\rm QFT}=\infty$. However, we do not assume that these particles are free, because they can still interact by exchanging particles of spin 0,1, and 2. In this case, the second term in equation (\ref{eq:GGhhGrowth}) cannot restore causality since there are only finite number of exchanges. In fact, all HS exchanges in the second term can only make the causality violation worse as we increase $s$. Thus, causality requires that $\Lambda_{\rm gr} \lesssim m_{\rm min}$ even if all the HS particles satisfy $|g_J|\gtrsim \frac{m_J}{M_{\rm pl}}$. This implies that there is no QFT description exists in this case.

\subsubsection*{$\mathbf {D>4}$}
In $D>4$ dimensions, we can repeat the argument of section \ref{sec:UniqueJJhVertex} which now implies \cite{Afkhami-Jeddi:2018apj} 
\be
\langle G_J(p_1,z_1)G_J(p_3,z_3)h(q,z)\rangle_{\rm causal} =0 + \frac{1}{M_{\rm pl}}\O\(\frac{m_J}{\Lambda_{\rm gr}}\)\ .
\ee
This can only be consistent with the soft-theorem for all $G_J$ if  
\be
\Lambda_{\rm gr} \lesssim m_{\rm min}\ .
\ee

\subsubsection*{Gravitational decay of HS particles}
Finally, let us comment on our assumption about the gravitational decay (\ref{eq:decay}) of $G_J$ particles. It should be noted that this assumption can be relaxed if we include other sectors. If we relax the decay condition (\ref{eq:decay}), HS particles in the non-gravitational sector can  contribute to the gravitational phase-shift. For example, higher spin composites can decay to two gravitons with $|\lambda_G|\gg 1$. Our argument still applies as long as $|\lambda \lambda_G| \ll 1$, where $\lambda\sim g_J$ is the interaction strength as defined in (\ref{eq:interaction}). On the other hand, the argument of section \ref{sec:UniqueJJhVertex} breaks down for $|\lambda_G \lambda| \sim  \O(1)$. Nevertheless,  for $|\lambda \lambda_G| \ge 1$ we can replace external gravitons by another particle which belongs to a different sector that interacts with $G_J$ only gravitationally. For example, we can replace the external graviton by a spectating scalar $\psi$ and can make an identical argument with leads to the same final conclusion.\footnote{By ``spectating" scalar we mean that the scattering amplitude for $\psi G_J \rightarrow \psi G_J$ has t-channel poles only at locations corresponding to particles in the gravity sector. } 

~\\

Taken literally, these results imply that if our universe contained metastable HS  particles, causality would require new HS particles in the gravity sector much below the Planck scale. Furthermore, causality necessarily requires an infinite tower of HS particles in the gravity sector with unbounded spin \cite{Camanho:2014apa,Afkhami-Jeddi:2018apj,Caron-Huot:2016icg}.  In the next section we will argue that $\Lambda_{\rm gr}$ should be regarded as the string scale.

\subsection{Curved Spacetime}
The bounds of this section are obtained by studying local high energy scattering which should not be sensitive to the spacetime curvature. Hence, we expect that the same bounds hold even in curved spacetime as long as curvature is not too large. Moreover, causality in CFT$_d$ imposes rigorous constraints on the interactions of particles in AdS$_{d+1}$ \cite{Hofman:2008ar,Hartman:2015lfa,Afkhami-Jeddi:2016ntf,Afkhami-Jeddi:2017rmx,Afkhami-Jeddi:2018own,Afkhami-Jeddi:2018apj,Kologlu:2019bco}.  In particular, stable, elementary particles of spin $J \geq 3$ and mass $m_J$ cannot couple to AdS$_{d+1}$ gravity with $d \geq 3$  unless there exists an infinite tower of HS states with increasing $J$ \cite{Afkhami-Jeddi:2018apj}.  The tower of new states must begin near the mass scale $m_J$.  We recently discussed   the application of causality constraints to composite particles \cite{Kaplan:2019soo} which now implies that metastable HS particles in AdS can only be coupled to stringy gravity 
\be
\Lambda_{\rm gr} \lesssim m_{\rm min}\ .
\ee
 Metastability in AdS corresponds to operator mixing in the CFT, as the unstable particle will mix with its decay products \cite{Kaplan:2019soo}.  Hence, the above bound is reliable only when all three-point interactions are small compared to the AdS radius. 
 
Similar bounds should apply in de Sitter as well. Moreover, it is possible that causality imposes stronger constraints in de Sitter. After all, there is a tension between the Higuchi bound  \cite{Higuchi:1986py, Deser:2001us} and HS excitations in string theory along the Regge trajectory \cite{Noumi:2019ohm,Lust:2019lmq}.  It is known that HS particles produce characteristic signatures on certain inflationary observables \cite{Arkani-Hamed:2015bza, Kehagias:2017cym,Franciolini:2018eno,Bordin:2018pca, Arkani-Hamed:2018kmz}. So, a rigorous causality analysis in de Sitter, as discussed in \cite{Afkhami-Jeddi:2018apj},   will have an immediate application in inflation.

\section{Asymptotic Uniqueness and Emergence of Strings}\label{sec:CKSZ}

In the last section we have shown that the gravity sector in HS theories necessarily contains the graviton and an infinite tower of fine-tuned HS particles, where $\Lambda_{\rm gr}$ is the mass of the lightest HS particle in the gravity spectrum. Theories of weakly interacting HS particles are strongly constrained by S-matrix consistency conditions. We now follow \cite{Caron-Huot:2016icg} to explore the asymptotic structure of gravitational scattering amplitudes when the gravity sector contains HS states.

\subsection{CKSZ Uniqueness Theorem}
Causality, as discussed in section \ref{sub:causality}, implies that a four-point amplitude where a finite number of HS particles are exchanged is inconsistent with causality, because the exchange of a particle with spin $J$ produces a phase shift $\delta\sim s^{J-1}$. Hence, causality necessarily requires that four-point amplitudes can only have exchanges of an infinite tower of HS particles with increasing spin.

\begin{figure}
\begin{center}
\usetikzlibrary{decorations.markings}    
\usetikzlibrary{decorations.markings}    
\begin{tikzpicture}[baseline=-3pt,scale=0.4]

\begin{scope}[very thick,shift={(4,0)}]

\draw (0,0) circle [radius=3];
\draw[-latex]  (2.12,2.12)--(5,5) ;
\draw[-latex]  (-5,-5)--(-2.12,-2.12);
\draw [-latex] (5,-5)--(2.12,-2.12) ;
\draw [-latex](-2.12,2.12)--(-5,5) ;
\draw(3,3)node[above]{\large $p_4$};
\draw(-3,3)node[above]{\large $p_3$};
\draw(3,-3)node[below]{\large $p_2$};
\draw(-3,-3)node[below]{\large $p_1$};

\draw(5,5)node[above]{\large $\psi$};
\draw(-5,5)node[above]{\large $\psi^\dagger$};
\draw(5,-6.5)node[above]{\large $\psi^\dagger$};
\draw(-5,-6.5)node[above]{\large $\psi$};
\end{scope}
\end{tikzpicture}
\end{center}
\caption{\label{cksz} \small $2\rightarrow 2 $ scattering of $\psi$ particles.}
\end{figure}
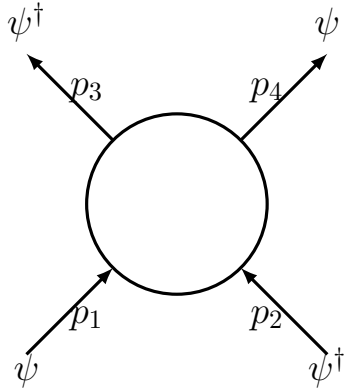

What are the possible theories with HS exchanges that obey the S-matrix consistency conditions of section \ref{sub:causality} and \ref{sub:SMatrix}? A version of this question has been addressed by the CKSZ theorem \cite{Caron-Huot:2016icg}. Consider   a $2\rightarrow 2 $  scattering of a scalar $\psi$ particle, as shown in figure \ref{cksz}, which involves the exchange of HS states. Any such amplitude at high energies must have the following properties \cite{Caron-Huot:2016icg}:\footnote{For a review see appendix \ref{appendix:cksz}.}
\begin{enumerate}
\item {The leading Regge trajectory is asymptotically linear.}
\item{The leading contribution to the inelastic part of the amplitude in the large $s$ and large impact parameter limit has a universal stringy form (\ref{inelastic}).}
\item{The scattering amplitude in the unphysical regime $s,t\gg 1$ coincides with the tree-level Gross-Mende amplitude (\ref{app_asym}).}
\item{The spectrum of the theory contains an infinite set of asymptotically parallel linear Regge trajectories.}
\end{enumerate}

Next we invoke  the CKSZ theorem to conclude that any weakly coupled UV completion of a theory of stable or metastable HS particles coupled to gravity must have a  gravity sector that contains an infinitely many asymptotically linear parallel Regge trajectories such that gravitational scattering amplitudes in the unphysical regime $s,t\gg \Lambda_{\rm gr}^2$ coincide with the tree-level closed string amplitude. This is perfectly consistent with the observation that infinite towers of HS particles in string theory lead to a well behaved S-matrix \cite{DAppollonio:2015fly,Amati:1987wq,Amati:1987uf,Amati:1988tn,Amati:1992zb,Amati:1993tb}. 

Let us emphasize that the CKSZ theorem applies only when massive higher-spin states are exchanged. However, in general a theory may contain HS particles that are finely tuned such that they are not exchanged in the scattering process \ref{cksz} or in any other $2\rightarrow 2$ scattering. This happens naturally, for example,  when HS particles are  charged under some global symmetry such as $\mathbb{Z}_2$. The CKSZ theorem does not say anything about this scenario. On the other hand, the argument of the preceding section implies that any theory with even one approximately elementary massive HS particle of mass $m_{J}$, however finely tuned, cannot be coupled to the graviton while preserving causality unless there exist other HS states $\sum X$ in the gravity sector at or below $m_{J}$. In other words, even if there is a massive HS particle that is not exchanged in any $2\rightarrow 2$ scattering process, its mere presence requires the gravity sector to include HS states $\sum X$, as shown in figure \ref{tower}, at $\Lambda_{\rm gr}\ll M_{\rm pl}$. Hence even in this scenario, the gravitational scattering of $\psi$ particles (figure \ref{cksz}) does include massive HS exchanges. Now the CKSZ theorem immediately implies that  any weakly coupled UV completion of the resulting theory  must have an asymptotically unique stringy gravity sector.

\subsection{Theory of HS Particles Coupled to Gravity}
In the last section, we argued that a consistent theory of metastable HS particles can be coupled to gravity while preserving causality  if and only if  the graviton exchanged scattering amplitude Reggeizes for $|t|\sim \Lambda_{\rm gr}^2$, where causality imposes an upper bound on $\Lambda_{\rm gr}$. Specifically,  if there is one HS particle in the $\{G_J\}$-sector which violates the weak gravity condition $|g_J| \le \frac{m_J}{M_{\rm pl}}$, we showed that $\Lambda_{\rm gr}\lesssim m_J$. 

\begin{figure}
\begin{center}
\usetikzlibrary{decorations.markings}    
\usetikzlibrary{decorations.markings}    
\begin{tikzpicture}[baseline=-3pt,scale=0.75]
\begin{scope}[very thick,shift={(4,0)}]
\draw[very thick,-latex]  (0,0) -- (-1.1,1.1);
\draw[very thick]  (-1,1) -- (-2,2);
\draw[very thick]  (-1.1,-1.1) -- (0,0);
\draw[very thick, -latex]  (-2,-2) -- (-1,-1);
\draw  (0,0) circle (1.5pt) ;
\draw [domain=0:3.5, samples=500]  plot (\x, {0.1* sin(5*pi*\x r)});
\draw(-2,2)node[above]{$\psi^\dagger$};
\draw(-2,-2)node[below]{$\psi$};
\draw(2,0.3)node[above]{ $h_{\mu\nu}+ \sum X$};
\end{scope}
\end{tikzpicture}
\end{center}
\caption{\label{tower} \small The tower of HS particles $\sum X$ must contribute to all gravitational interactions above $\Lambda_{\rm gr}$.}
\end{figure}
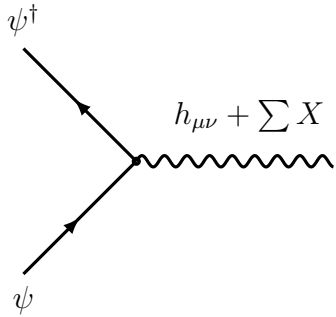

Now we consider a $2\rightarrow 2 $  gravitational scattering of the spectating scalar $\psi$.\footnote{Note that in the preceding discussion, we assumed that $\psi$ is charged under some global $U(1)$ symmetry. This is just a simplifying assumption which is not essential for the final conclusion. The global $U(1)$ symmetry implies that the u-channel resonances are absent. However, even if we have had the u-channel, the same argument holds. So, the particle $\psi$ does not have to be charged for the CKSZ theorem to be applicable.}   The important point is that the graviton exchange is accompanied by the tower of HS particles $\sum X$, as shown in figure \ref{tower}. The CKSZ theorem enables us to make the following conclusions about the gravity sector of the full theory. 

{\bf Low energy limit:} First consider the physical regime $s>0$ and $t<0$ of the gravitational scattering amplitude. Specifically, for small angle Regge scattering $|s|\rightarrow \infty$ we can  write 
\be\label{eq:lowenergy}
\lim_{|s|\gg |t|} A_{\text gravity}(s,t) =F(t) (-s)^{j(t)}\ .
\ee
For $|t|\ll \Lambda_{\rm gr}^2$, this should recover the result for a single graviton exchange imposing 
\be
\lim_{|t|\ll \Lambda_{\rm gr}^2} F(t)=\frac{1}{M_{\rm pl}^2 t}\ , \qquad \lim_{|t|\ll \Lambda_{\rm gr}^2}j(t)=2\ 
\ee
for $t<0$. Moreover, unitarity imposes $\lim_{|t|\ll \Lambda_{\rm gr}^2}j'(t)>0$.

{\bf Leading Regge trajectory:} The CKSZ theorem implies that the leading Regge trajectory must be asymptotically linear\footnote{In contrast to appendix \ref{appendix:cksz}, we are using $\frac{\alpha'}{2}$ as the Regge slope to be consistent with string theory conventions. } 
\be\label{regge_qcd}
j(t)=2+\frac{\alpha'}{2} t+\delta j(t)\ ,
\ee
where $\delta j(t)$ has the property that $\delta j(0)=0$ and $\delta j(t\gg 1)\ll \alpha' t/2$. The fact that $j(0)=2$ follows from the requirement that the spectrum of exchanged particles must include gravitons. Note that masses of heavy spinning particles on the leading Regge trajectory can be approximated by
\be
m_J^2 \approx \frac{2J}{\alpha'}\ .
\ee
Hence, we can identify 
\be
\alpha' \sim \frac{1}{\Lambda_{\rm gr}^2}\ .
\ee

\begin{figure}[h]
\begin{center}
\includegraphics[width=0.4\textwidth]{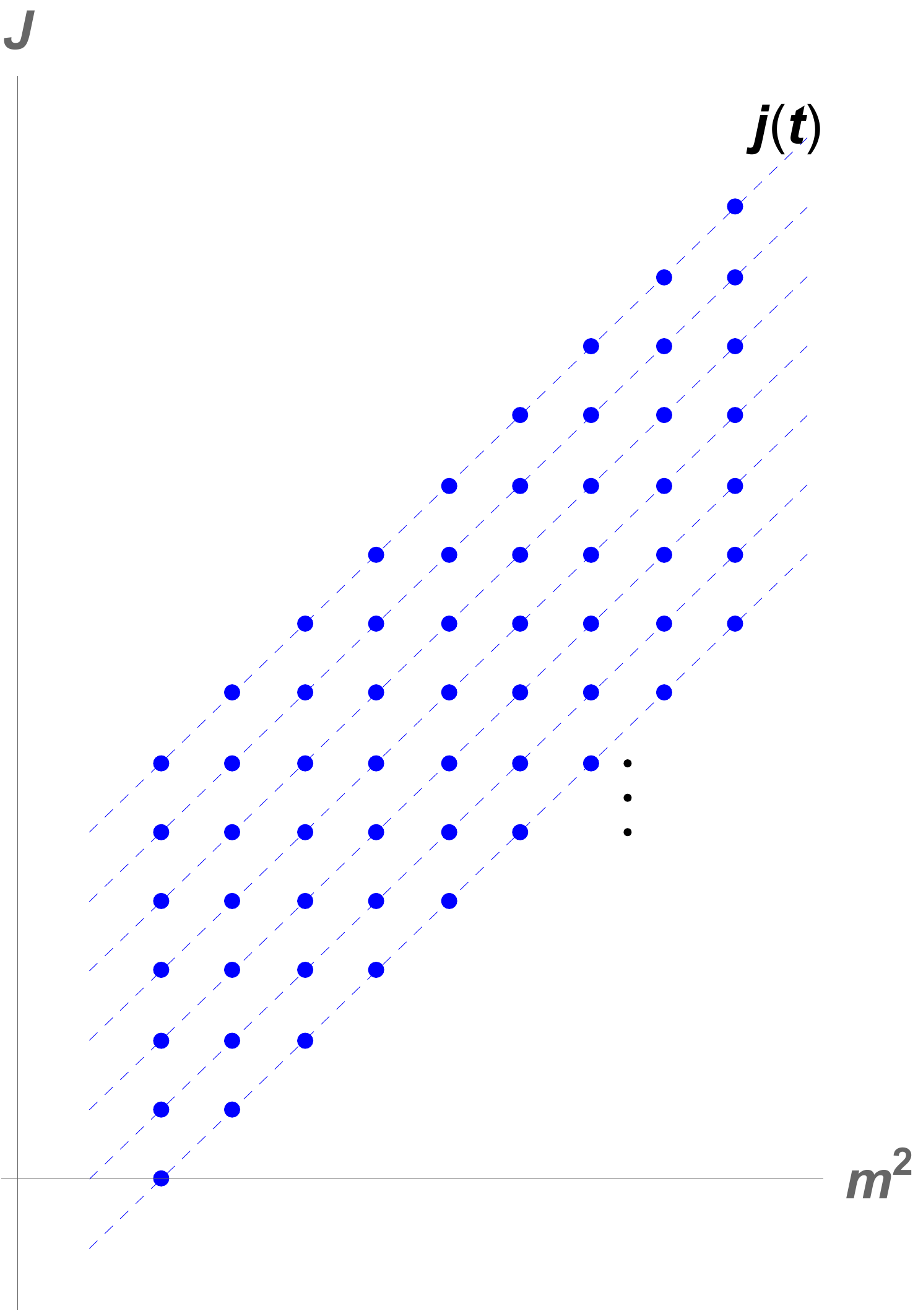}
\end{center}
\caption{ \small  We couple a theory of metastable HS particles to gravity. This is a schematic plot of the spectrum of particles with large spins that the gravity sector must contain in the resulting theory. Blue dots represent actual particles and dashed blue lines are Regge trajectories, where the origin of the axes is arbitrary. Consistency of the S-matrix requires that the Regge trajectories must be asymptotically linear, parallel, and equispaced. Note that  there are an infinite tower of  particles for any fixed spin. \label{fig:spectrum}}
\end{figure}

{\bf High energy amplitude:} The S-matrix consistency conditions also require the existence of an infinite tower of asymptotically parallel Regge trajectories. This can be alternatively stated in the following way. The amplitude in the  regime $s,t\gg 1$ must coincide with the tree-level Gross-Mende string amplitude\footnote{Note that the large $t,s$ limit is defined by $\mbox{Re}\ t, s\gg 1$ with $\mbox{Im}\ t, s>0$ such that poles are avoided.} 
\be\label{amp_qcd}
\lim_{s,t\gg 1}  A_{\text gravity}(s,t)=A_0 \exp\(\frac{\alpha'}{2}\((s+t)\ln (s+t)-s\ln s-t\ln t\)\)
\ee
with $\alpha' \sim \frac{1}{\Lambda_{\rm gr}^2}$. This amplitude in the  regime $|s|\gg |t| \gg 1$ can be rewritten as a sum over an infinite set of asymptotically linear, parallel, and equispaced Regge trajectories 
\be
\lim_{\alpha' s\gg \alpha' t\gg 1}  A_{\text gravity}(s,t)=F(t)\sum_{n=0}^\infty \frac{(\alpha' t)^n}{2^{2n} (1)_n} s^{\frac{\alpha'}{2} t-n}\ ,
\ee
where $F(t)=A_0 e^{\alpha't/2}t^{-\alpha' t/2}$. This implies that the gravity sector must have an infinite tower of heavy particles for any fixed spin. In general, for arbitrarily large spin $J$ the spectrum contains particles with masses $m(J)^2\approx 2\Lambda_{\rm gr}^2(J+n)$ for all non-negative integer $n$, as shown in figure \ref{fig:spectrum}.   

{\bf Large impact parameter:} The scattering amplitude in the high energy $s\gg \Lambda_{\rm gr}^2$ and large impact parameter $b\gg \sqrt{\log(s/\Lambda_{\rm gr}^2)}/\Lambda_{\rm gr}$ regime is completely universal. In this limit, the real part of the amplitude in the impact parameter space is determined by the graviton exchange 
\be
\mbox{Re}\ A_{\text gravity}(b,s) \approx \frac{s^2}{M_{\rm pl}^2}\log\(\frac{b}{L}\)\ ,
\ee 
where $L$ is the IR cut-off. On the other hand, the  asymptotic linearity of the leading Regge trajectory implies that the leading contribution to the inelastic part of the amplitude is universal  as well
\be\label{eq:inelastic}
\mbox{Im}\ A_{\text gravity}(b,s) \approx e^{-\frac{b^2}{2\alpha' \log(\alpha' s/2)}}
\ee
with $\alpha' \sim \frac{1}{\Lambda_{\rm gr}^2}$.

{\bf Bound on $\Lambda_{\rm gr}$:} In a generic theory $\Lambda_{\rm gr}$ is arbitrary and can be as large as $M_{\rm pl}$. However, causality imposes strong restrictions on $\Lambda_{\rm gr}$ for HS theories. In particular, the causality constraints of the previous section implies that for a theory of metastable HS particles 
\be\label{final_bound}
\Lambda_{\rm gr}\lesssim \mbox{Min}[\Lambda_{\rm gr}^{(J)}, J\ge 3]\ ,
\ee
where 
\begin{align}\label{eq:weakgravity}
&\Lambda_{\rm gr}^{(J)}  = m_J \(\frac{|g_J| M_{\rm pl}}{ m_J}\)^{\gamma(J)}  \qquad ~~ |g_J|\gtrsim \frac{m_J}{M_{\rm pl}}\nonumber\\
&\Lambda_{\rm gr}^{(J)} = m_J\qquad ~~~~~~~~~~~~~~~~~~~~~~|g_J|\lesssim \frac{m_J}{M_{\rm pl}}
\end{align}
and $\gamma(J)$ is defined in (\ref{gamma}).

{\bf Emergence of strings:} The fact that the inelastic part of the amplitude (\ref{eq:inelastic}) is universal and non-zero has important implication. This amplitude suggests the theory contains extended objects of  size $\sqrt{\log(s/\Lambda_{\rm gr}^2)}/\Lambda_{\rm gr}$. So, it is natural to identify 
\be
\Lambda_{\rm gr}\approx M_{\rm string}\ .
\ee
Indeed, the $\sqrt{\log(s)}$ enhancement of the size is exactly what is expected from quantization of the strings \cite{Karliner:1988hd,Susskind:1993aa}. Furthermore, the spectrum of particles at large spin, as shown in figure \ref{fig:spectrum}, coincides with the spectrum of particles in tree-level string theory. Since, the tower of HS particles $\sum X$ accompanies the graviton, we should interpret them as excitations of a fundamental closed string.  

{String scattering amplitudes, for large $t$ and $s$, are truly short distance phenomena. However, we still have some computational control because  at high energies strings are stretched over large lengths and hence we can ignore string oscillations. This simplifies the computation greatly making it possible to calculate the exact high energy behavior of string amplitudes at each order in perturbation theory \cite{Gross:1987kza}. Moreover, the leading high energy behavior of string amplitudes is independent of the exact quantum numbers of scattering particles.\footnote{For a review see appendix \ref{app:GM}.} The asymptotic amplitude (\ref{amp_qcd}), of course, coincides with the large $s, t$ limit of the Virasoro-Shapiro amplitude as expected. }

Therefore, we conclude that the bound (\ref{final_bound}) should be interpreted as an upper bound on the string scale. 

\subsection{A Weak Gravity Condition}
Let us now consider a theory of stable or metastable HS particles coupled to gravity in 4d. The resulting theory, as we have shown, must contain stringy states above $\Lambda_{\rm gr}$. However, we can still obtain a low energy QFT description for a set of light HS particles by integrating out states above $\Lambda_{\rm gr}$. Hence, a QFT description exists for a HS particle of mass $m_J$ and interaction strength $g_J$ only when $\Lambda_{\rm gr}\gg m_J$. From equation (\ref{eq:weakgravity}), we see that a parametric separation between $\Lambda_{\rm gr}$ and  $m_J$ necessarily requires 
\be\label{eq:last}
 |g_J|\gtrsim \frac{m_J}{M_{\rm pl}}\ .
\ee
This is precisely the statement that the gravitational interaction between the particle is weaker than the non-gravitational interaction. On the other hand,  for $ |g_J|\lesssim \frac{m_J}{M_{\rm pl}}$ we have $\Lambda_{\rm gr} \lesssim m_J$ and hence such a HS particle only has a stringy description. Thus, for a traditional QFT description HS particles must obey the weak gravity condition (\ref{eq:last}). Equivalently, all metastable HS particles with masses $m_J\ll \Lambda_{\rm gr}$ in 4d must obey the weak gravity condition (\ref{eq:wg}) which states that the gravitational part of the $2\rightarrow 2$ scattering amplitude must be smaller than the non-gravitational part in the impact parameter space for $\Lambda_{\rm gr}\gg \sqrt{s}\gg \frac{1}{b}, m_J$. 

The fact that there can be a parametric separation between $\Lambda_{\rm gr}$ and $m_J$ even for $ |g_J|\gtrsim \frac{m_J}{M_{\rm pl}}$ is true only in 4d. In this sense, 4d is special because it  allows for a field theoretic approximation of HS particles coupled to gravity.

Finally, we wish to emphasize that the weak gravity condition is only a necessary condition. For example, when we couple a theory of finite number of elementary HS particles to gravity, the resulting theory must obey $\Lambda_{\rm gr} \lesssim m_{\rm min}$. Hence, this theory does not have a QFT description even if all the HS particles satisfy the weak gravity condition (\ref{eq:last}) or (\ref{eq:wg}). This implies that a free massive HS particle can only be coupled to a gravity theory which is stringy.


\section*{Acknowledgements}
It is our pleasure to thank Nima Afkhami-Jeddi, Simon Caron-Huot, Liam Fitzpatrick, Shamit Kachru, David Kaplan,  Ami Katz, Juan Maldacena, and Amirhossein Tajdini  for several helpful discussions.  We were supported in part by the Simons Collaboration Grant on the Non-Perturbative Bootstrap. JK was supported in part by NSF grant PHY-1454083.


\appendix

\section{Phase-Shifts and Time-Delays}\label{appendix:decay}
In this appendix we present a more physical argument that implies the positivity of the phase-shift. First, we ensure that we are in the weakly coupled regime by imposing $\delta \ll 1$. We now replace the particle 1 by a coherent state of particles with a fixed polarization. Moreover, because of the weak coupling, we can take the mean occupation number to be large without making $\delta$ large. Bose enhancement then ensures that the polarization of particle 3 is complex conjugate of that of particle 1. The tree-level phase-shift now can be naturally exponentiated by studying the propagation of the particle $1$ in a background with $\mathcal{N}\sim1/|\delta|$ independent shockwaves, each of which is   created by a particle 2 in a fixed coherent state \cite{Camanho:2014apa} (for a pictorial representation see  figure \ref{nshock}). Of course, this approximation is valid only in the weakly coupled regime where the scattering processes are independent events. Moreover, the argument is more subtle when the particle 1 has a finite size. A careful analysis \cite{Camanho:2014apa,Kaplan:2019soo} ensures that this set-up is reliable in the regime  $\frac{m_1 r_1}{b^2 s}\ll |\delta| \ll 1$, where $m_1$ and $r_1$ are mass and radius of the particle 1, respectively. One can alway satisfy this condition when $\delta$ grows with $s$ implying exponentiation of the tree-level phase-shift. Hence, in this set-up the tree-level phase-shift $\delta$, when grows with $s$, determines the time-delay and must be non-negative .

\subsection*{Decay Rates and Time-Delays}

\begin{figure}
\centering
\includegraphics[scale=0.30]{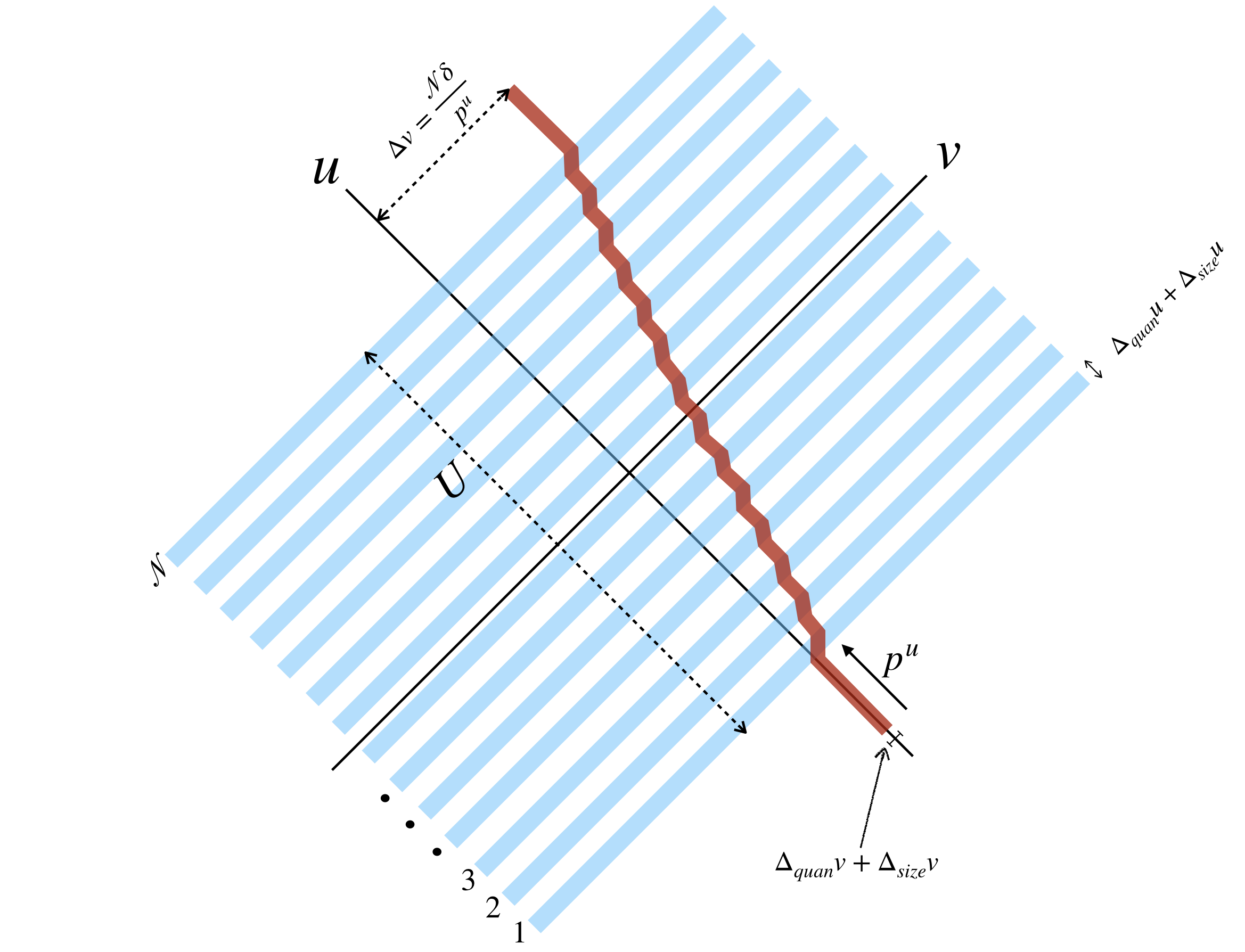}
\caption{\small Total time delay for a coherent state of incoming $G_J$ particles after crossing $\N$ independent shockwaves sourced by another particle ($G_0/h_{\mu\nu}/\psi$) can be large enough to violate asymptotic causality. }\label{nshock}
\end{figure}

In general, one might worry that a metastable HS particle can decay before we can detect any time-advance in our eikonal scattering set-up. Here we argue that the eikonal scattering set-up of \cite{Kaplan:2019soo} remains reliable even when HS particles have finite lifetime. We closely follow the argument of section 5.2 of \cite{Kaplan:2019soo}. We consider the scenario in which a $G_J$ particle of spin $J$  traveling in a shockwave sourced by the another particle which can be a $G_0$ or a graviton $h_{\mu\nu}$ or a spectating scalar $\psi$ which interacts with the $\{G_J\}$-sector via gravity. At tree-level, the amplitude is $1+i \delta$, where $|\delta| \ll 1$ in order for the theory to be weakly coupled. The Shapiro time delay of the particle $G_J$ is related to the phase shift 
\be
\Delta v= \frac{\delta}{p^u}\ ,
\ee
where, $p^u>0$ is the $u$-component of the momentum of particle $G_J$ (see figure \ref{nshock}). The tree-level approximation is reliable only when $|\Delta v| p^u \ll 1$. This tree-level effect can be amplified by performing the thought experiment of \cite{Camanho:2014apa}  in which a coherent state (with large occupation number) of  particle $G_J$ is propagating   in a background with $\mathcal{N}$ independent shockwaves  created by $G_0/h_{\mu\nu}/\psi$ particles.  Bose enhancement ensures that the incoming and outgoing states are exactly the same implying that the phase shift is the same for each of these $\N$-processes. In the limit $\delta \rightarrow 0$ and $\mathcal{N}\rightarrow \infty$ with $\mathcal{N}\delta$ fixed, the total amplitude is $(1+i \delta)^\mathcal{N}\approx e^{i \mathcal{N} \delta}$, implying that the total phase-shift is $\mathcal{N} \delta$.  For $G_J$, a causality violation can only be detected if and only if $ |\Delta v|$ is larger than all uncertainties associated with the thought experiment. This requires $\N |\delta| > m_J r$, where $r$ is the size of the particle $G_J$ with mass $m_J$  \cite{Kaplan:2019soo}.

The $G_0/h_{\mu\nu}/\psi$ particles with momentum $P^v$ that create shocks can only be localized over a distance
\be
\Delta_{\text{quan}} u \sim \frac{1}{P^v}\ .
\ee
Therefore, we can only get $\mathcal{N}$ independent shocks if  the entire process of scattering through $\mathcal{N}$ shocks takes null time
\be
U=\frac{\mathcal{N}}{P^v}\ .
\ee
The argument of \cite{Kaplan:2019soo} then imposes
\be
\label{grrm}
b^2 p^u>U> \frac{m_J r}{P^v |\delta|}\ .
\ee
The LHS inequality follows because we do not want the wavefunctions of the particles to spread by more than the impact parameter during the process.  The RHS inequality follows from an inequality $\N |\delta| > m_J r$ which demands a delay larger than the size of the particle in units of its Compton wavelength.

The causality argument breaks down if the decay time (in the lab frame) of the HS particle $t_{decay} < U$, where 
\be
t_{decay} = \(\frac{p^u}{m_J} \) t_{\rm com}\ .
\ee
Decay time of $G_J$ in the center of mass frame is $t_{\rm com}$. The particle $G_J$ is highly boosted and the factor of $p^u/m_J$ takes into account the requisite time dilation. We conclude that our set-up is reliable if  
\be
t_{\rm com}> \frac{m_J^2r }{ |\delta| s}\ .
\ee
Thus, we can trust our set-up by ensuring   
\be
s>\frac{m_J^2r }{ |\delta| t_{\rm com}}
\ee
without making $\delta$ large. This can always be satisfied in the regime $s\gg m_J^2, 1/b^2$ when $\delta$ grows with $s$, provided $t_{\rm com}m_J$ is not vanishingly small. Of course, this is reasonable because even unstable particles can travel arbitrarily large distances when they are sufficiently boosted.

\subsection*{Regime of Validity for the Interference Set-Up}
The interference bound requires the incoming state 1 and the outgoing state 3 to be a linear combination of two different particles. In general, these particles can have different masses and hence different momenta. So, if we wait for a long time, two different incoming particles will move away from each other. Thus we can trust our interference bound only if the null time  $U$ for the scattering process is not very large. Let us now make this  more precise.

First note that the incoming state 1 and the outgoing state 3 have the following momenta for $\vec{q}=0$
\be
p^\mu=\(p^u,\frac{m^2}{p^u},\vec{0} \)
\ee
when $p^u\gg m$. So, in null time $U$ the particle moves along $v$-direction by
\be
\delta v= U \(\frac{m}{p^u}\)^2\ .
\ee 
Hence the scattered beam has a width which grows with $U$
\be
\delta v_{width}= U \frac{\Delta m^2}{\(p^u\)^2}\sim  U \(\frac{m_J}{p^u}\)^2
\ee
which should be thought of as an additional error of the interference experiment. Therefore, we should ensure that 
\be
\frac{\N |\delta|}{p^u} >  \delta v_{width} \approx \frac{\N}{P^v}  \(\frac{m_J}{p^u}\)^2\ .
\ee
Moreover, we must also be in the weakly coupled regime. Therefore, the interference experiment is reliable only if
\be\label{validity2}
 1\gg |\delta|  \gg \frac{1}{s b^2}\ , \frac{m_J^2}{s}\ .
\ee
This can always be achieved in our interference set-up as long as $s\gg m_J^2, \frac{1}{b^2}$.

\section{Scattering Kinematics}
\label{app:ScatteringDetails}

In the eikonal limit, the momentum of particles are parametrized as follows\footnote{The metric is given by (\ref{eq:metric}). We use the convention: $A^\mu=(A^u,A^v,\vec{A}).$}
\begin{align}\label{kin4}
&p_1^\mu = \left(p^u,\frac{1}{p^u} \left(\frac{\vec{q} \;^2}{4}+m_1^2\right),  \frac{\vec{q}}{2} \right)\ , \qquad p_3^\mu =  \left(\bar{p}^u,\frac{1}{\bar{p}^u} \left(\frac{\vec{q} \;^2}{4}+m_3^2\right),  -\frac{\vec{q}}{2} \right)\ , \nonumber\\
& p_2^\mu= \left( \frac{1}{P^v}\left(\frac{\vec{q} \;^2}{4}+m_2^2\right), P^v,-\frac{\vec{q}}{2} \right)\ , \qquad p_4^\mu= \left( \frac{1}{\bar{P}^v}\left(\frac{\vec{q} \;^2}{4}+m_4^2\right),\bar{P}^v, \frac{\vec{q}}{2} \right)\ ,
\end{align}
where, $p^u,\bar{p}^u,P^v,\bar{P}^v>0$ and  $p_1^\mu- p_3^\mu\equiv q^\mu$ is the transferred momentum of the exchange particle. The eikonal limit is defined as $p^u, P^v \gg |q|,m_i$. In this limit  $p^u\approx \bar{p}^u, P^v \approx \bar{P}^v$  and the Mandelstam variable $s$ is  given by $s= - (p_1+p_2)^2 \approx  p^u P^v$. Massless particles have only transverse polarizations but massive particles can have both transverse and longitudinal polarizations. General polarization tensors can be constructed using the following transverse and longitudinal polarization vectors
\begin{align}\label{vectors}
&\epsilon^\mu_{T,\lambda}(p_1) = \left(0, \frac{\vec{q}\cdot \vec{e}_\lambda^{\;(1)}}{ p^u},  \vec{e}_\lambda^{\;(1)} \right)\ , \qquad \epsilon_L^\mu(p_1) = \left(\frac{p^u}{m_1},\frac{1}{ m_1 p^u} \left(\frac{\vec{q}\;^2}{4} - m_1^2 \right),  \frac{\vec{q}}{2m_1} \right)\ , \nonumber\\
& \epsilon^\mu_{T,\lambda}(p_3) = \left( 0,-\frac{\vec{q}\cdot \vec{e}_\lambda^{\; (3)}}{ p^u},  \vec{e}_\lambda^{\; (3)} \right)\ , \quad \epsilon_L^\mu(p_3) = \left(\frac{p^u}{m_3},\frac{1}{ m_3 p^u} \left(\frac{\vec{q}\;^2}{4} - m_3^2 \right),  - \frac{\vec{q}}{2m_3} \right) \ ,
\end{align}
where vectors $e_\lambda^{\mu} \equiv (0,0, \vec{e}_{\lambda})$ are complete orthonormal basis in the transverse direction $\vec{x}_\perp$. We can define $\epsilon^\mu_{T,\lambda}(p_2), \epsilon^\mu_{T,\lambda}(p_4)$ and $\epsilon_L^\mu(p_2), \epsilon_L^\mu(p_4)$ in a similar way. We will use the following null polarization vectors for external gravitons (when applicable)
\begin{align}
&\epsilon_h^\mu(p_1) =\frac{1}{\sqrt{2}}\( \epsilon^\mu_{T, \hat{x}}(p_1)-i \epsilon^\mu_{T, \hat{y}}(p_1)\)\ , \qquad \epsilon_h^\mu(p_3) = \frac{1}{\sqrt{2}}\(\epsilon^\mu_{T, \hat{x}}(p_3)+i \epsilon^\mu_{T, \hat{y}}(p_3)\)\ ,\nonumber\\
&\epsilon_h^\mu(p_2) = \frac{1}{\sqrt{2}}\(\epsilon^\mu_{T, \hat{x}}(p_2)-i \epsilon^\mu_{T, \hat{y}}(p_2)\)\ , \qquad \epsilon_h^\mu(p_4) =\frac{1}{\sqrt{2}}\( \epsilon^\mu_{T, \hat{x}}(p_4)+i \epsilon^\mu_{T, \hat{y}}(p_4)\)\ ,
\end{align}
where $\hat{x}=(0,0,1,0)$ and $\hat{y}=(0,0,0,1)$. Similarly, the external HS particle $G_J$ has the following polarizations (when applicable)
\begin{align}
&\epsilon_J^\mu(p_1) =\frac{1}{\sqrt{2}}\(i \epsilon^\mu_{L}(p_1) +\epsilon^\mu_{T, \hat{x}}(p_1)\)\ , \qquad \epsilon_J^\mu(p_3) =\frac{1}{\sqrt{2}}\(-i \epsilon^\mu_{L}(p_3) +\epsilon^\mu_{T, \hat{x}}(p_3)\)\ ,\nonumber\\
&\epsilon_J^\mu(p_2) =\frac{1}{\sqrt{2}}\(i \epsilon^\mu_{L}(p_2) +\epsilon^\mu_{T, \hat{x}}(p_2)\)\ , \qquad \epsilon_J^\mu(p_4) =\frac{1}{\sqrt{2}}\(-i \epsilon^\mu_{L}(p_4) +\epsilon^\mu_{T, \hat{x}}(p_4)\)\ .
\end{align}

\section{The Soft Theorem and Graviton Induced Mixing}\label{appendix:soft}
The bound (\ref{bound2}) played a significant role in our main argument. So, it is of value to understand how the bound (\ref{bound2}) follows from the soft theorem.  First of all, the expansion (\ref{expansion}) must be consistent with the soft theorem. We will follow the elegant formalism of \cite{Laddha:2017ygw} to relate $\Gamma_{GG'h}$ and $G G'$ mixing without a graviton in the soft limit $q\rightarrow 0$. 

Let us define $G_\alpha$ where $\alpha$ runs over all states of the $\{G_J\}$-sector. To be specific, $\{G_\alpha\}$ belongs to some large reducible representation of the local Lorentz group. The most general quadratic 1PI effective action of $G_\alpha$ in 4d (without gravity) is given by
\be\label{action_GG}
S_{GG}=\frac{(2\pi)^4}{2}\int d^4 p_1 d^4 p_3 G_\alpha(p_1) {\mathcal K}^{\alpha\beta}(p_3) G_{\beta}(p_3) \delta^4(p_1+p_3)
\ee
with the convention ${\mathcal K}^{\alpha\beta}(p)={\mathcal K}^{\beta\alpha}(-p)$. Since, there is no kinetic mixing 
\be
{\mathcal K}^{\alpha\beta}(p)=0 \qquad \alpha\neq \beta\ .
\ee
We now couple a soft graviton to a pair of finite energy $G_J$ particles. This can be done by covariantizing the action (\ref{action_GG}) following \cite{Laddha:2017ygw}. The action for one soft (on-shell) graviton and two $G_J$ particles is uniquely fixed by gauge invariance up to order $q$. In particular, the leading gauge invariant action in the soft limit is given by \cite{Laddha:2017ygw}
\begin{align}\label{sen}
S_{GGh}=-\frac{(2\pi)^4}{2}\int d^4 p_1 d^4 p_3 \delta^4(p_1+p_3+q) h_{\mu\nu} p_3^\nu G_\alpha(p_1)\frac{\p  {\mathcal K}^{\alpha\beta}(p_3)}{\p p_{3\mu}}G_\beta(p_3)+ \O(q)\ , 
\end{align} 
The higher order corrections can be found in \cite{Laddha:2017ygw}, however, these corrections are not important for our argument. The important fact is that the $q^0$ part of the on-shell three-point function $\Gamma_{GG'h}$ must be proportional to off-diagonal elements of ${\mathcal K}^{\alpha\beta}$. Hence,

\be
\Gamma^{(0)}=0
\ee
in the expansion (\ref{expansion}). When combined with the causality constraint (\ref{eq:intf_bound}) this implies  (\ref{bound2}). Our final conclusion remains unchanged even when the kinetic mixing between different $G_J$'s is non-zero but small. 

\section{Summary of the CKSZ Uniqueness Theorem}\label{appendix:cksz}

Any weakly coupled theory in which HS particles are exchanges is strongly constrained by S-matrix consistency conditions \cite{Caron-Huot:2016icg}. In particular, these theories exhibit universal asymptotic behavior which follows from the CKSZ uniqueness theorem \cite{Caron-Huot:2016icg}. In this appendix, we summarize the CKSZ theorem. 

Let us consider   a $2\rightarrow 2 $  scattering of the $\psi$ particles as shown in figure \ref{cksz}. Note that the physical regime is $t<0$ and $s>4m_\psi^2$. We assume that $\psi$ is charged under some global $U(1)$ symmetry. This is just a simplifying assumption which is not essential for the final result. The global $U(1)$ symmetry implies that the u-channel resonances are absent.

We make the assumption that the scattering amplitudes obey the S-matrix conditions discussed in section \ref{sub:SMatrix}.  We now discuss these assumptions in more detail. In any weakly coupled theory, scattering amplitude $A(s,t)$ is a meromorphic function with only simple poles at the location of resonances $\{m_i^2\}$
\be\label{amp}
\lim_{s\rightarrow m_i^2} A(s,t) \rightarrow \frac{1}{s-m_i^2} \sum_{n,L}f_{n,L}^2 P_L\(1+\frac{2t}{m_i^2-4m_\psi^2}\)\ ,
\ee
where $P_L$ is the Legendre polynomial.\footnote{Discussion of this section applies in any spacetime dimensions. For $D\neq4$, the Legendre polynomials should be replaced by the Gegenbauer polynomials.} Furthermore, unitarity requires that $f_{n,L}^2\ge 0$ in (\ref{amp}). The amplitude is also crossing symmetric $A(s,t)=A(t,s)$. We intend to explore $|t|,|s| \rightarrow \infty$ limit of the scattering amplitude. To be specific, by $|t|,|s| \rightarrow \infty$ we mean that $|t|,|s|$   are large enough such that all intermediate scales  decouple. Hence, in this limit it is expected that  the expression (\ref{eq:regge}) holds even when $|t|/|s|$ is small but fixed. In addition, we assume that there is no accumulation point in the spectrum implying there are finite number of particles in the spectrum with masses below any mass scale.

Any physical theory is expected to be causal at least at low energies. Hence, the constraint (\ref{eq:causal}) requires that the Regge amplitude (\ref{eq:regge}) must obey \be\label{app:causal}
\lim_{|s|\gg |t|, m_i^2} |A(s,t)| \lesssim s^2
\ee
for $t=m_0^2$, where $m_0$ is the mass of the lightest particle on the leading Regge trajectory. Besides, for $t<m_0^2$ but not far below $m_0^2$, the Regge amplitude must grow slower than $s^2$ implying $\lim_{ |s|\rightarrow \infty} s^{-2} A(s,t)=0$ in this regime.\footnote{In \cite{Caron-Huot:2016icg}, it was assumed that there exists a particle in the spectrum with spin $L$ such that 
\be
\lim_{ |s|\rightarrow \infty} s^{-L} A(s,t)=0
\ee
for some fixed $t$.  We do not make this assumption since causality implies the above condition for $L=2$ and $t\le m_0^2$. 
}

The CKSZ theorem partially answers the question: what are the possible theories with HS exchanges that obey the above S-matrix consistency conditions? Any such theory at high energies exhibits some universal behaviors. In particular, any such theory must be a theory of strings with a spectrum containing an infinite set of asymptotically parallel linear Regge trajectories. 

\subsubsection*{Linearity of the leading Regge trajectory}
The above consistency conditions impose that the leading Regge trajectory must be asymptotically linear \cite{Caron-Huot:2016icg}
\be\label{regge_asym}
j(t)=\alpha' t+\cdots\ , \qquad t\gg 1
\ee
where $\alpha'$ is the slope of the trajectory which is not constrained by the S-matrix consistency conditions. Roughly speaking, $\alpha'\sim m^{-2}$ where $m$ is the mass of a typical light HS particle.

\subsubsection*{Existence of strings}
By performing a saddle point approximation, one can relate the scattering amplitude in the high energy $\alpha' s\gg 1$ and large impact parameter $b^2\gg \alpha'$ regime with $j(t\gg 1)$.  The  asymptotic linearity of the leading Regge trajectory immediately implies that the leading contribution to the inelastic part of the amplitude in the impact parameter space is universal  \cite{Caron-Huot:2016icg}
\be\label{inelastic}
\mbox{Im}\ A(b,s) \approx e^{-\frac{b^2}{4\alpha' \log(\alpha' s)}}\ .
\ee
This amplitude suggests the existence of extended objects in the theory. The typical size of these extended objects is roughly $\sqrt{\alpha' \log(\alpha' s)}$ which is consistent with string theory.

\subsubsection*{Asymptotic uniqueness of the amplitude}
Actually the S-matrix consistency condition imposes a stronger restriction on the high energy amplitude. The amplitude in the unphysical regime $s,t\gg 1$ must coincide with the tree-level Gross-Mende amplitude \cite{Caron-Huot:2016icg}
\be\label{app_asym}
\lim_{s,t\gg 1}  A(s,t)=A_0 \exp\(\alpha'\((s+t)\ln (s+t)-s\ln s-t\ln t\)\)
\ee
which coincides with the large $s, t$ limit of both closed string and open string four-point amplitudes. This amplitude cannot be reproduced just by the leading Regge trajectory (\ref{regge_asym}). In fact, (\ref{app_asym}) requires the existence of an infinite tower of asymptotically parallel, linear, and equispaced Regge trajectories. 

The S-matrix consistency conditions do impose constraints on the subleading correction of both (\ref{regge_asym}) and (\ref{app_asym}).  In fact, under some additional assumptions, it was argued in \cite{Sever:2017ylk} that these subleading corrections are universal as well. The subleading corrections can be interpreted as the massive ends correction of relativistic strings.

\section{String Scattering Amplitudes at High Energies}\label{app:GM}
String scattering amplitudes have universal soft behavior at high energies. In a classic paper, Gross and Mende showed that the exact leading behavior of string scattering amplitudes for large $t$ and $s$ can be computed, order by order in perturbation theory, in a systematic way by using a saddle-point approximation \cite{Gross:1987kza}. Here we review the tree-level Gross-Mende amplitude.

\subsection*{Closed String Amplitude}
From the structure of string amplitudes, it is clear that the leading high energy behavior is independent of the exact quantum numbers of scattering particles. This allows us to  consider scattering of massless string states without loss of generality.  We start with the four-scalar tree-level closed string amplitude
\be\label{string}
A_{\rm string}^{\rm closed}=\frac{g_s^2}{\mbox{Vol}(SL(2;C))} \int \mathcal{D} X^\mu e^{-\frac{1}{2\pi \alpha'}\int d^2z\ \p X\cdot \bar{\p} X} \prod_{i=1}^4 V_i
\ee
where $z=\sigma+ i \tau$ and vertex operators are defined in the usual way
\be
V_i= \int d^2{z_i} e^{i p_i^\mu X_{\mu}(z_i, \bar{z}_i)}\ .
\ee
Note that the normalization factor $\mbox{Vol}(SL(2;C))$ is there to remind us that there is a residual $SL(2;C)$ symmetry that acts on $(z_i, \bar{z}_i)$. For simplicity, we are assuming that all particles have incoming momentum, such that $\sum_i p^\mu_i=0$. We define $s$, $t$, and $u$ such that our final result is consistent with the convention of the rest of the paper
\be
s=-(p_1+p_2)^2\approx -2p_1\cdot p_2\ , \quad t=-(p_1+p_3)^2\approx -2p_1\cdot p_3\ , \quad u=-(p_1+p_4)^2\approx -2p_1\cdot p_4\ .
\ee

The amplitude (\ref{string}) can be evaluated exactly for specific external states. For example, when external states are tachyons, then (\ref{string}) leads to the Virasoro-Shapiro amplitude. The tree level amplitude (\ref{string}) can be evaluated exactly even when external states have spins. Of course, in that case the final amplitude is a more complicated function of $s$, $t$, $u$ and polarizations. However, we are only interested in the  large $t$ and $s$ limit in which the amplitude is insensitive to the details of asymptotic states. 

Since all momentum transfers are large, we can now perform a saddle point approximation in (\ref{string}). The saddle point is  the solution of the Laplace equation
\be
\bar{\p}\p X^\mu(z,\bz)=-i\pi \alpha' \sum_{i=1}^4 p_i^\mu\delta(z-z_i)\delta(\bz-\bz_i)
\ee
which has the solution
\be
X^\mu(z,\bz)=-\frac{i\alpha' }{2}\sum_{i=1}^4 p_i^\mu \ln |z-z_i|^2\ .
\ee
Using this solution, we can approximate (\ref{string})
\be
A_{\rm string}^{\rm closed}\approx \frac{g_s^2}{\mbox{Vol}(SL(2;C))} \int \prod_{i=1}^4  d^2{z_i} \prod_{i,j=1}^4 |z_i-z_j|^{\frac{\alpha'}{2}p_i\cdot p_j}\ .
\ee
We now make use of the $SL(2;C)$ symmetry by choosing 
\be
z_1=z\ , \qquad z_2=0\ , \qquad z_3=1\ , \qquad z_4=\infty
\ee
and similarly for $\bz_i$. This yields
\be
A_{\rm string}^{\rm closed}\approx g_s^2 \int d^2{z} |z|^{\alpha' p_1 \cdot p_2}|1-z|^{\alpha' p_1\cdot p_3}=g_s^2 \int d^2{z} |z|^{-\alpha' s/2}|1-z|^{-\alpha' t/2}\ .
\ee
The above integral converges only for $\alpha' s, \alpha' t< 2$ and $\alpha'(s+t)> 2$. However, for any $s$ and $t$ there is a unique analytic continuation in terms of gamma functions. For large $|s|, |t|$ the analytically continued amplitude can be obtained by performing a saddle point approximation which yields  
\be
z=\bz=\frac{s}{s+t}\ .
\ee
Finally, we obtain the high energy closed string scattering amplitude
\be\label{stringy}
A_{\rm string}^{\rm closed}\approx g_s^2 \exp\(-\frac{\alpha'}{2}\(s\ln |s|+t\ln |t|-(s+t)\ln |s+t|\)\)\ .
\ee
Note that the final answer is independent of the theory and external states. Of course, the exact prefactor will depend on these details. 

The above amplitude has some interesting properties that we now review. Let us first compare the above result with generic QFT amplitudes. There is a robust lower bound on high energy amplitudes given by Cerulus and Martin \cite{Cerulus:1964cjb}. The bound says that for large $s$ and $t$ with $t/s$ fixed
\be
|A(s,t)| \ge \exp\(- C(t/s) \sqrt{s}\ln s\)\ ,
\ee
where $C(t/s)>0$. The Cerulus-Martin bound assumes unitarity, existence of a finite mass gap and locality. Locality is imposed by requiring that the amplitude is polynomially bounded in $s$ for large $s$ and fixed $t$. Clearly, the string amplitude (\ref{stringy}) violates the Cerulus-Martin bound. It is believed that the violation of the locality condition is responsible for the violation of the Cerulus-Martin bound in string theory. The remarkable fact about the amplitude (\ref{stringy}) is that it is local enough to be asymptotically causal,\footnote{For discussions about causality in string theory see \cite{Erler:2004hv,Martinec:1993zv}.} however, sufficiently non-local to violate the Cerulus-Martin bound.

\subsection*{Open String Amplitude}
A similar analysis can be performed for open string four-point amplitudes in large $|s|, |t|$ limit. In that case, one obtains
\be
A_{\rm string}^{\rm open}\approx g_s \exp\(-\alpha'\(s\ln |s|+t\ln |t|-(s+t)\ln |s+t|\)\)\ 
\ee
which is the asymptotic limit of the Veneziano amplitude.

\end{spacing}

\bibliographystyle{utphys} 
\bibliography{CausalityBib}

\end{document}